\documentclass[%
 aip,
 amsmath,amssymb,
 reprint,%
]{revtex4-1}

\usepackage{graphicx}
\usepackage{dcolumn}
\usepackage{bm}

\usepackage[utf8]{inputenc}
\usepackage[T1]{fontenc}
\usepackage{mathptmx}
\usepackage{subfigure}

\begin{document}

\preprint{AIP/123-QED}

\title[Floating Potential of Spherical Dust in Collisionless Magnetised Plasmas]{Floating potential of spherical dust in collisionless magnetised plasmas}

\author{L. M. Simons}
\email{ls5115@ic.ac.uk}
\affiliation{
Blackett Laboratory, Imperial College, London SW7 2AZ United Kingdom
}%
\author{M. Coppins}%
\affiliation{
Blackett Laboratory, Imperial College, London SW7 2AZ United Kingdom
}%

\date{\today}

\begin{abstract}
Determining the equilibrium charge of conducting spheres in plasmas is important for interpreting Langmuir probe measurements, plasma surface interactions and dust particle behaviour. The Monte Carlo code Dust in Magnetised Plasmas (DiMPl) has been developed for the purpose of determining the forces and charging behaviour of conducting spheroids under a variety of conditions and benchmarked against previous numerical results. The floating potentials of spheres in isothermal, collisionless, hydrogen plasmas as a function of magnetic field strength and size relative to Debye length are studied using DiMPl and compared with new results from the N-body tree code (pot) and recent particle in cell measurements. The results of all three simulations are similar, identifying a small range at modest ion magnetisation parameters over which the electron current is reduced relative to the ion current. The potential as a function of magnetic field strength is found to be relatively insensitive to dust size for dust smaller than the Debye length. The potential of large dust is found to depend less strongly on flow speed for modest magnetic field strengths and to decrease with increasing flow speed in the presence of strong magnetic fields for smaller dust. A semi-empirical model for the potential of small dust in a collisionless plasma as a function of magnetic field strength is developed which reproduces the expected currents and potentials in the high and low magnetic field limit.
\end{abstract}

\maketitle

A long-standing theoretical problem in plasma physics is the determination of the floating potential of a conducting object immersed in a plasma~\cite{Mott-Smith1926}. Plasma-surface interactions are of crucial importance in space plasmas due to the ubiquitous dust component~\cite{Mendis1994}. These theories also underpin the interpretation of Langmuir probe diagnostics which are routinely used in making measurements of the plasma density and temperature~\cite{Hopkins1986}. This essential feature of plasmas is fundamental to our understanding of solid and liquid particle dynamics in plasmas~\cite{Spitzer1941,Bacharis2010}, spacecraft~\cite{Garrett1981,Olson2010} and dust transport in the scrape off layer in tokamaks~\cite{Stangeby1984}. 


Controlling the accumulation and mobilization of dust in magnetic confinement fusion devices is acutely important for safely achieving the goal of stable power production~\cite{Malizia2016}. Principally, the accumulation of high Z impurities in the core plasma, which is enhanced by tungsten dust ablation, must be avoided due to the strong bremstrahhlung losses~\cite{Putterich2010} and the potential to cause disruptions~\cite{DeVries2011}. Dust inventory in tokamaks presents a biological hazard due to the toxicity of materials used such as beryllium~\cite{Longhurst2004}, with their small size creating the possibility for inhalation in a loss of vacuum event. The activity of machines must also be minimised which is particularly difficult with tungsten plasma facing components and carbon dust which can easily capture tritium due to it's chemical reactivity~\cite{Roth2008}.

Fundamental to understanding dust motion in plasmas is modelling their equilibrium charge. To attain tractable analytic expressions, approximations must be introduced for the magnetic or electric field, the plasma Debye length, $\lambda_{d}$, and the plasma drift velocity relevant to the regime of interest. The elementary model for predicting the equilibrium potential is the Orbital Motion Limited (OML) theory, which assumes a sphere of radius $a_{d}$ in a collisionless, plasma with no drift or external electromagnetic field in the limit $a_{d}\ll\lambda_{d}$~\cite{Mott-Smith1926,Allen1992}. In the opposite case, $a_{d}\gg\lambda_{d}$, for very large dust relative to Debye length, a Modified (OML) theory (MOML) is used~\cite{Willis2012}. These can be trivially extended to include an arbitrary ion flow velocity by assuming ions a large distance away have a shifted Maxwellian velocity distribution~\cite{Kimura1998}, theories known as Shifted OML (SOML) and Shifted MOML (SMOML)~\cite{Willis2012}. The SOML theory in particular is frequently used as a benchmark for kinetic and fluid plasma codes investigating the floating potential of solid surfaces~\cite{Hutchinson2005,Delzanno2013,Rizopoulou2013}. An alternative solution for the potential distribution surrounding the dust can be acquired through the orbital motion approach~\cite{Bohm1949,Bernstein1959,Laframboise1966,Allen1992}. This provides estimates for the floating surface potential of dust grain of any size and is often used for validation. For dust in contact with a high temperature plasma such as is found in a tokamak, the charging is complicated further by electron emission processes~\cite{Pigarov2005,Holgate2018} through thermal~\cite{Dushman1923} and secondary electron emission~\cite{Wooldridge1957}, ion impact~\cite{Thomas1992} and photon absorption~\cite{Schmidt-Ott1981} which can contribute significantly to the charging process\cite{Rizopoulou2018}. A combination of SOML and a number of these electron emission processes constitutes the standard charging model employed in most simulations of dust transport in laboratory and tokamak plasmas~\cite{Pigarov2005, Krasheninnikov2011, Bacharis2012, Autricque2016, Tanaka2007, Gervasini2017, Vignitchouk2014,Rizopoulou2018}. In all cases, the influence of the magnetic field on charging for the purpose of tracking motion is neglected despite the fact that the magnitude of magnetic fields present in many of these environments significantly alter the charging characteristics of dust~\cite{Vignitchouk2017,Laframboise1976, Patacchini2007, Lange2016, Thomas2016}.

With the introduction of a uniform magnetic field, analytic solutions for the equilibrium surface potential of dust exist only in specific limits~\cite{Sonmor1991}. For an uncharged sphere, partial analytic solutions for the floating potential have been derived~\cite{Whipple1965} in the limit $a_{d}\ll\lambda_{d}$. For charged spheres, analytic solutions for the upper and lower bounds on the floating potential have been calculated in the limit where $a_{d}\ll\lambda_{d}$ ~\cite{Sanmartin1970,Laframboise1976} and $a_{d}\gg\lambda_{d}$~\cite{Sonmor1991}. 

For the case of $a_{d}\ll\lambda_{d}$, the currents of ions and electrons are most simply conceptualised through the critical fields which are reached when the thermal gyro-radii of each plasma species becomes small relative to the dust grain size~\cite{Tsytovich2003}. For magnetic field strengths where the electron gyro-radius, $\rho_{\perp,e}$, is comparable to the dust grain size, the motion of electrons becomes restricted perpendicular to field lines and their current is reduced whilst the ions current remains relatively unaffected. At larger magnetic field strengths, the ions motion also becomes restricted as the ion gyro-radius, $\rho_{\perp,i}$, decreases, leading to an increase in the potential. For intermediate magnetic field strengths, the potential still must either be calculated through numerical integration of the kinetic equations~\cite{Sonmor1991} or estimated based on results of particle-in-cell simulations~\cite{Patacchini2007,Lange2016}. 
Developing a closed form expression for the floating potential which can be solved directly is therefore desirable to alleviate the computationally intensive task of it's numerical solution. Recent progress has been made in this regard with the development of an analytical solution for large dust in the scrape-off layer in tokamaks~\cite{Vignitchouk2017}, when $\rho_{\perp,e}\ll a_{d}\leq \rho_{\perp,i}$ and $a_{d}\gg\lambda_{d}$. However this only provides a solution over a narrow range of particle sizes and plasma conditions.

In this paper, the results of the Dust in Magnetised Plasmas (DiMPl) code are presented and used to develop a semi-empirical model for calculating the floating potential of conducting spheres with $a_{d}\leq\lambda_{d}$ in collisionless, hydrogen plasmas. In section \ref{sec:Dust in Magnetised Plasmas (DiMPl)}, the DiMPl Monte Carlo code is presented and benchmarked against previous numerical solutions. In section \ref{sec:Simulation results}, the results of DiMPl are presented, studying the effect of varying magnetic field strength, dust size and flow velocity on floating potential. The results of these simulations are used in section \ref{sec:Semi-empirical formulation of floating potential} to constrain the theoretically motivated semi-empirical model for the floating surface potential of conducting dust grains in magnetic fields.
\section{\label{sec:Dust in Magnetised Plasmas (DiMPl)}Dust in Magnetised Plasmas (DiMPl)}
\begin{figure}[h]
\centering
\includegraphics[clip,width=0.5\textwidth]{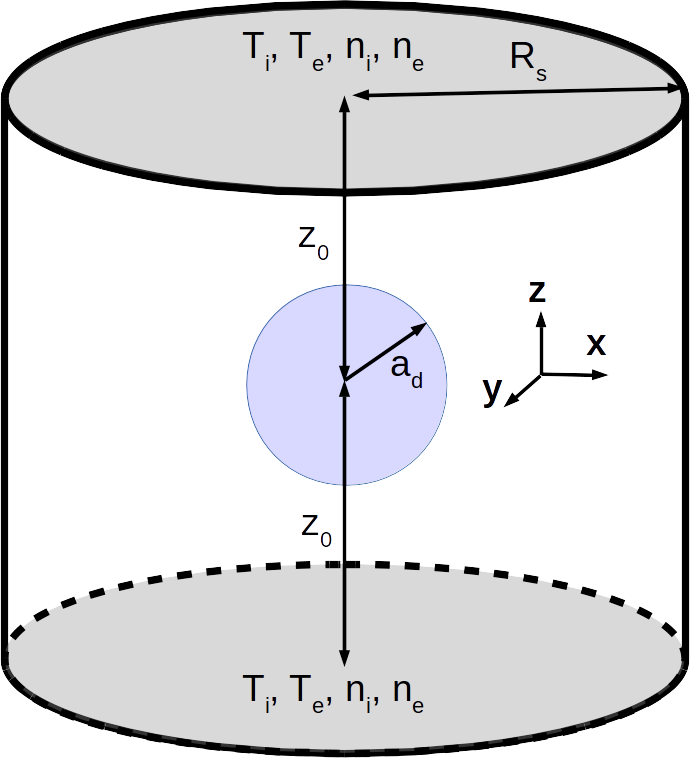}
\caption{Schematic of simulation domain in cartesian coordinates bounded by circular surfaces at $\pm z_{0}a_{d}$ of radius $R_{s}$ and a central spherical surface of radius $a_{d}$. The distant plasma has a well defined ione and electron temperatures $T_{i}$, $T_{e}$ and densities $n_{i}\sim n_{e}=n_{0}$.}
\label{Fig:SimulationSchematic}
\end{figure}

The Dust in Magnetised Plasma (DiMPl) Code is a Monte-Carlo code that tracks particles continuously through a cylindrical geometry with a spherically symmetric electric and constant, uniform magnetic field without considering inter-particle interactions. A schematic for the simulation domain is shown in figure \ref{Fig:SimulationSchematic}. The sphere is assumed to be a perfectly conducting spheroid of charge $Q_{d}$ centered on the origin of the simulation domain. Simulations assume a collisionless, ion and electron plasma flowing with a velocity parallel to the magnetic field in the $\widehat{\underline{z}}$ direction having a collisional length scale much larger than all other length scales.

\subsection{\label{subsec:Boundary Conditions}Boundary Conditions}
A cylindrical simulation boundary is the most natural choice since the spherical symmetry of the problem is broken by the uniform linear magnetic field and flow. Particles are injected uniformly over two circular planes perpendicular to the magnetic field, at a height $\pm z_{0}a_{d}$ from the origin. The infinite planes defined by $z=\pm z_{0}a_{d}$ and a spherical surface of radius $a_{d}$ centered around the origin define the simulation boundaries. 

\begin{figure*}[t]
\centering
\includegraphics[clip,width=\textwidth]{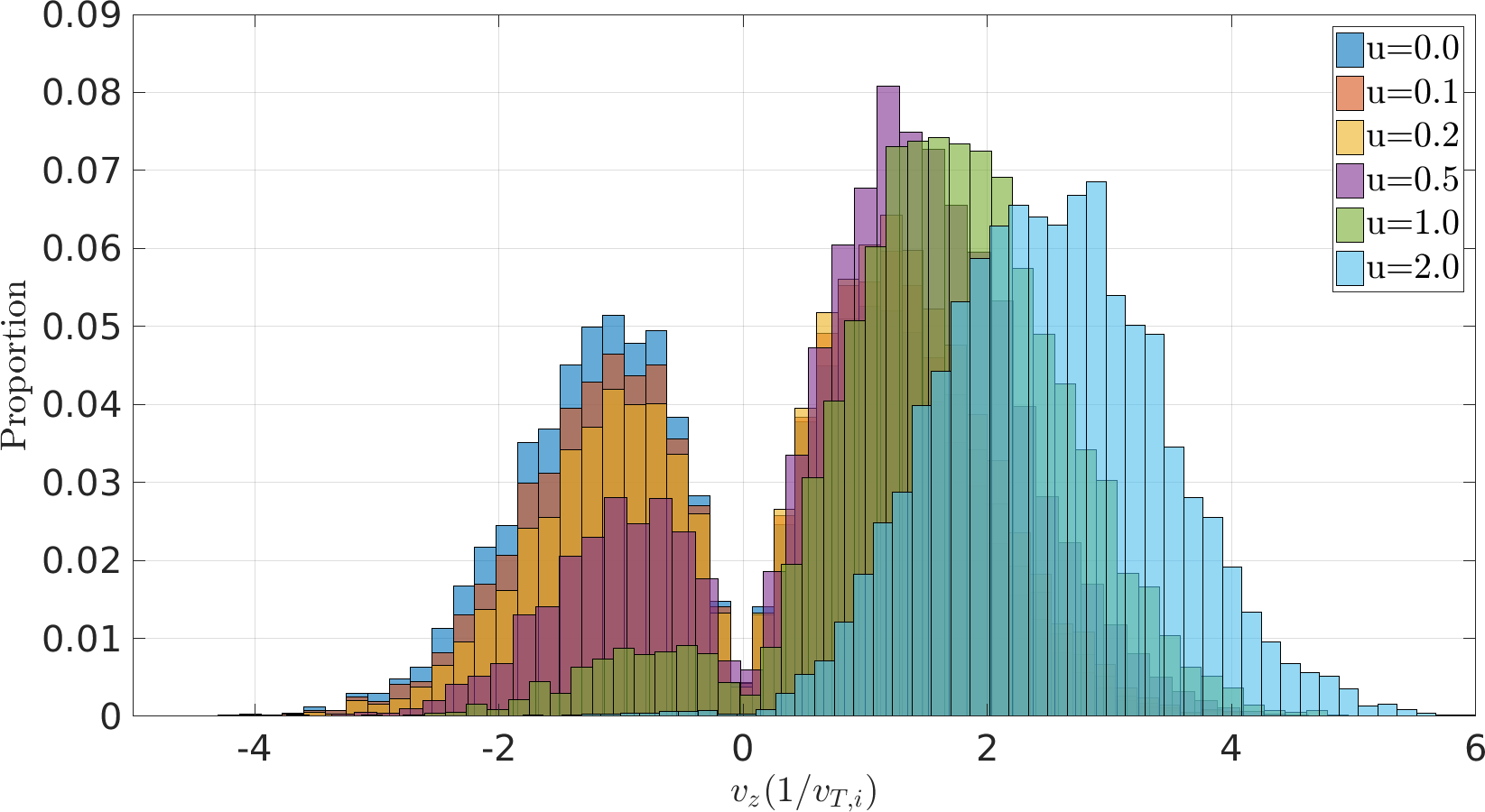}
\caption{Normalised velocity distribution of ions injected over the circular boundaries at $z=\pm z_{0}a_{d}$ for different normalised drift velocities $u=U/v_{T,i}$.}
\label{Fig:VelocityDistribution}
\end{figure*}

For sufficiently large $z_{0}$, the electrostatic potential at the injection surface is comparable to the plasma potential and the distribution is unaffected by the dust electric field. The surface is then assumed to be a thermalised, collisional source of ions and electrons denoted by subscript $s=i,e$, having well defined temperatures, $T_{s}$, masses $m_{s}$, thermal velocities, $v_{T,s}=\sqrt{k_{B}T_{s}/m_{s}}$, and equal ion and electron densities, $n_{i}$ and $n_{e}$ respectively, in accordance with quasi-neutrality $n_{i}\sim n_{e}=n_{0}$. The distribution of particle velocities parallel to the circular surface $f_{v}(v_{x},v_{y})$, with velocity components $v_{x}$ and $v_{y}$ in Cartersian coordinates, is approximated by a Maxwell-Boltzmann distribution 
\begin{equation}\label{eq:ParallelVelocityDistribution}
f_{v}(v_{x},v_{y})=\dfrac{m_{s}}{2\pi k_{B}T_{s}}e^{-\dfrac{m_{s}(v_{x}^{2}+v_{y}^{2})}{2k_{B}T_{s}}}.
\end{equation}
Perpendicular to the plane, the velocity distribution differs from this due to the preferential passage of particles with higher velocities through the plane. For this reason, the velocity probability distribution of $v_{z}$ is calculated from the 1-way flux, $\Gamma_{s}$, flowing at velocity $U$ parallel to the magnetic field following the method of Makkonen~\cite{Makkonen2015}
\begin{equation}\label{eq:OneWayVelocityDistribution}
f_{v}(v_{z})=\dfrac{n_{0}v_{z}}{\Gamma_{s}(U)}\sqrt{\dfrac{m_{s}}{2\pi k_{B}T_{s}}}e^{-\dfrac{m(v_{z}-U)^{2}}{2k_{B}T_{s}}},
\end{equation}
\begin{equation}\label{eq:OneWayFluxOfParticles}
\begin{split}
\Gamma_{s}(U)=&n_{0}\Big[\sqrt{\dfrac{k_{B}T_{s}}{2\pi m_{s}}}e^{-\dfrac{m_{s}U^{2}}{2k_{B}T_{s}}},\\
&+\dfrac{U}{2}\Big(1+erf\Big(U\sqrt{\dfrac{m_{s}}{2k_{B}T_{s}}}\Big)\Big)\Big]. \\
\end{split}
\end{equation}
Figure \ref{Fig:VelocityDistribution} shows example vertical velocity distributions for injected particles from both the upper and lower plane for different normalised drift velocities, $\underline{u}=U/v_{T,i}\widehat{\underline{z}}$.

When a magnetic field of strength $B$ is present, charged particles perform gyro-orbits around magnetic field lines, with a typical, thermal gyro-radius defined by $\rho_{T,s}=v_{T,s}m_{s}/eB$. Only particles with an initial radial position on the order of the gyro-radius are considered since the majority of particles with at a larger impact parameter will miss the sphere, incurring unnecessary computational expense. This is equivalent to truncating the integral over perpendicular velocity at a large value. For this reason, the circular injection planes for ions and electrons are defined with areas $A_{s}=\pi R_{s}^{2}$ and radii $R_{s}= a_{d}+b(\rho_{T,s}+\lambda_{d})$ where $b$ is the impact factor, a simulation parameter. The probability of a particular species being injected at a particular surface is calculated from the ratio of the flux of particles of that species through the surface to the total flux, given by equation (\ref{eq:OneWayFluxOfParticles}). The probability of injecting an ion, $P_{i}^{\pm}$, or electron, $P_{e}^{\pm}$, at position $\pm z_{0}a_{d}$, from a distribution given by equation \ref{eq:OneWayVelocityDistribution}, through surfaces of areas $A_{i}$ and $A_{e}$ respectively, is
\begin{equation}\label{eq:InjectionProbabilities}
\begin{split}
&P^{\pm}_{i}=\dfrac{\Gamma_{i}(\pm U)A_{i}}{A_{i}(\Gamma_{i}(U)+\Gamma_{i}(-U))+A_{e}(\Gamma_{e}(U)+\Gamma_{e}(-U))},\\
&P^{\pm}_{e}=\dfrac{\Gamma_{e}(\pm U)A_{e}}{A_{i}(\Gamma_{i}(U)+\Gamma_{i}(-U))+A_{e}(\Gamma_{e}(U)+\Gamma_{e}(-U))}.\\
\end{split}
\end{equation}
After specifying the initial conditions, ions and electrons are injected at random with probabilities given by equation (\ref{eq:InjectionProbabilities}). Particles are given an initial random velocity following the distributions in equation (\ref{eq:ParallelVelocityDistribution}) and equation (\ref{eq:OneWayVelocityDistribution}) and a random initial position uniformly distributed over the circular areas $A_{s}$.

\subsection{Solving Equations of Motion}
The motion of non-relativistic charged particles with position $\underline{r}(t)$ at time $t$ traversing a region with a central electric field $\underline{E}(\underline{r}(t))=E(\underline{r}(t))\widehat{\underline{r}}$ due to the charged spherical dust grain and a constant uniform magnetic field $\underline{B}=B\underline{\widehat{z}}$ is given by the Lorentz force law. Using the initial conditions $\underline{r}(t_{0})$ and $\underline{v}(t_{0})$ at $t=t_{0}$ provided by the boundary conditions, the position after a time step $\Delta t_{0}$ is calculated via numerical integration using the Boris algorithm~\cite{Boris1970}. 

In the limit of small dust where the Debye length is much greater than the size of the sphere $\lambda_{d}\gg a_{d}$, the potential of the sphere is unshielded and the ions and electrons experience a bare Coulomb electric field
\begin{equation}\label{eq:CoulombFields}
\underline{E}(\underline{r}(t))=\dfrac{Q_{d}}{4\pi\epsilon_{0}\underline{r}^{2}(t)}\widehat{\underline{r}},
\end{equation}
where $\epsilon_{0}$ is the permittivity of free space. For larger dust sizes, the dust charge becomes screened by the plasma and the electric field is approximated instead by a Debye-Huckel potential
\begin{equation}\label{eq:DebyeFields}
\begin{split}
\underline{E}(\underline{r}(t))=&\dfrac{Q_{d}}{4\pi\epsilon_{0}|\underline{r}(t)|}e^{-\dfrac{|\underline{r}(t)|-a_{d}}{\lambda_{d}}}\Big(\dfrac{1}{|\underline{r}(t)|}+\dfrac{1}{\lambda_{d}}\Big)\widehat{\underline{r}}, \\
\lambda_{d}=&\sqrt{\dfrac{\epsilon_{0}k_{B}T_{e}}{e^{2}n_{0}}},
\end{split}
\end{equation}
where $e$ is the fundamental charge of an electron. The presence of a magnetic field and flow can lead to non-spherically symmetric electrostatic potential distributions around a charged object. The imposition of these central electric fields is therefore only strictly valid when $|\underline{u}|\leq1$ and $a_{d}>\rho_{T,i}$. Magnetic field strengths are measured in terms of the magnetisation parameter, which is given by the ratio of the dust grain radius to mean gyro radius of species $s$ defined by $\beta_{s}\equiv a_{d}/\rho_{T,s}$. For typical conditions in the scrape off layer of a tokamak with $B=5T$ and $T=10eV$, the ion magnetisation parameter is $\beta_{i}\simeq0.02a_{d}(\mu m)$ for a hydrogen plasma. Dust sizes are characterised by the normalised Debye length $\tilde{\lambda_{d}}\equiv \lambda_{d}/a_{d}$.

The system of equations is solved explicitly until one of three conditions are satisfied. If at any point, $\vert\underline{r}(t)\vert\leq a_{d}$, then the particle is considered to be collected and contribute fully their charge and momentum to the dust grain. If a particle reaches a point where $\underline{r}(t)\cdot\underline{\widehat{z}}>z_{0}a_{d}$, the particle leaves the simulation domain and the lost momentum and charge are recorded. A maximum number of reflections parallel to the magnetic field is defined to improve the performance of the simulation. This prevents the excessive computational effort expended on tracking particles which become temporarily trapped in the attractive potential. Particles of the attracted species which exceed $15$ reflections are considered to return to infinity while for the repelled species, only one reflection is required. This limit has in previous work been shown to provide a good estimate of the currents to a sphere~\cite{Sonmor1991}.

\subsection{\label{Sec:Method}Method}
\begin{figure}[h]
\centering
\includegraphics[clip,width=0.5\textwidth]{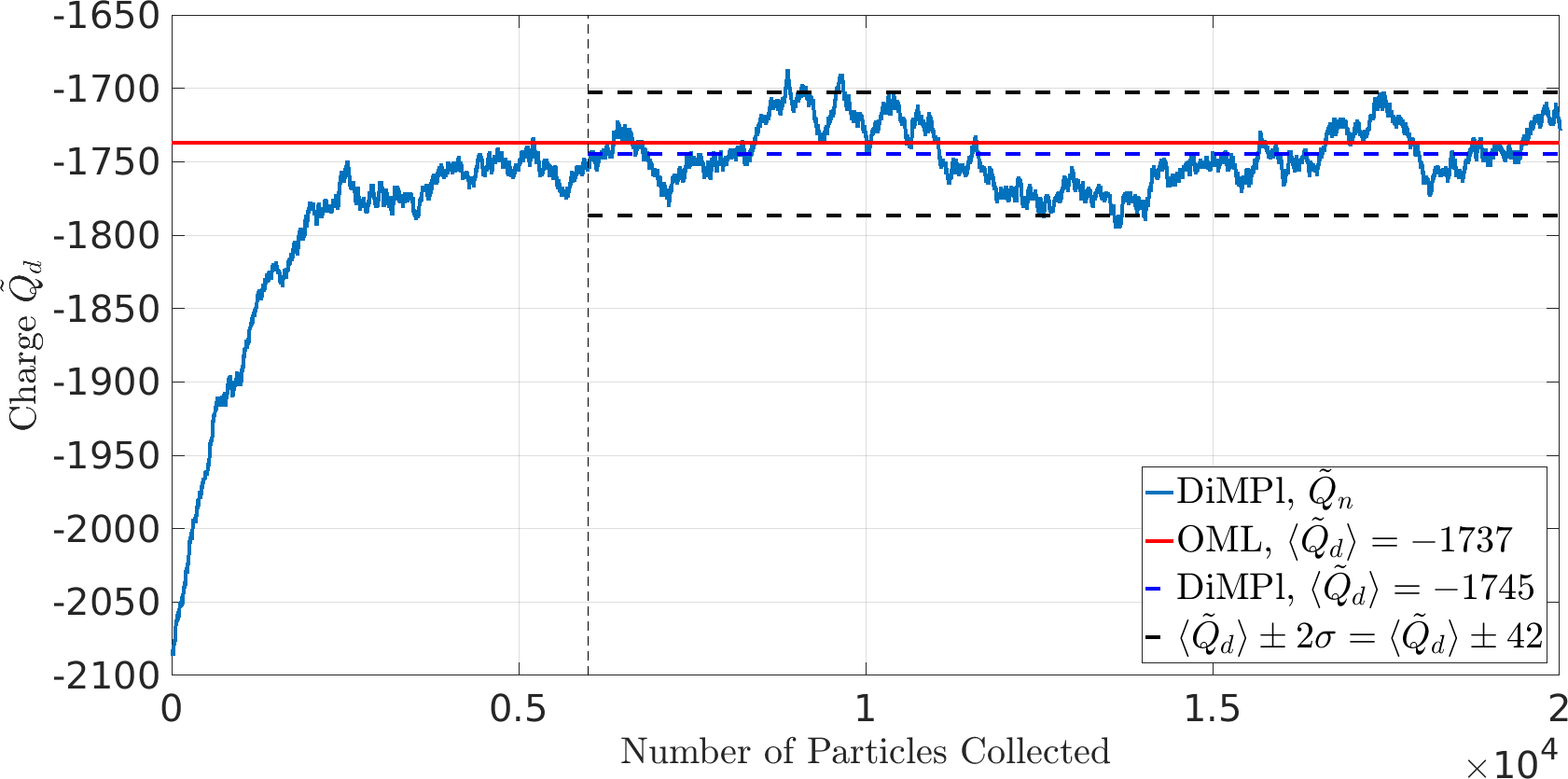}
\caption{Normalised charge, $\tilde{Q}_{d}$ , as a function of number of particles collected and equilibrium charge measurement $\langle \tilde{Q}_{d}\rangle$ with a self-consistent surface potential for $\beta_{i}=0.01$ compared with the theoretical OML value. A vertical dashed line at $6000$ collected particles indicates the start of the equilibrium measurement.}
\label{Fig:MethodBExample}
\end{figure}

Each ion or electron trajectory is solved independently in parallel without a global record of time for an isothermal hydrogen plasma with mass ratio $\mu=m_{i}/m_{e}$ where $m_{i}$ and $m_{e}$ are the hydrogen ion and electron masses respectively. The default simulation parameters used were $b=5$ and $z_{0}=50$ with time step $\Delta t_{0}=0.01eB/m_{e}$ with default plasma parameters $T_{i}=T_{e}=1eV$ and $n_{0}=10^{18}m^{-3}$ unless otherwise stated. When calculating the floating potential of the dust grain, the potential varies dynamically as charges are collected. After a transitory equilibration phase, a time independent measurement of the equilibrium charge can be made, as shown in figure \ref{Fig:MethodBExample} for $\beta_{i}=0.01$. The mean potential is calculated after reaching equilibrium, indicated by the vertical black line, from the average normalised charge $\langle\tilde{Q}\rangle$ where $Q=e\tilde{Q}$. The uncertainty on the mean decreases with increasing number of particles collected. The results form a non-Markovian time series since a given measurement of the equilibrium charge, $\tilde{Q}_{n}$, is correlated with subsequent and previous measurements of charge, affecting the interpretation of the standard errors. 

It may appear from figure \ref{Fig:MethodBExample} that a numerical solution of the time dependent equation $I(t)=dQ(t)/dt$ has been performed. However, because there is no global track of time and because the rate of particle collection depends upon the surface potential, the time dependence of the current to the dust grain cannot be determined. On the other hand, if the surface potential is fixed, the time averaged flux of particles at the dust surface will be constant, for a sufficiently large number of injected particles. By fixing the dust surface potential in analogy with a biased probe, this was exploited to measure the time evolution of quantities.

The time, $t_{G}$, to generate $G_{s}$ particles of a particular species is related to the constant flux at the outer boundary $\Gamma_{s}A_{s}=G_{s}/t_{G}$. Assuming that for a sufficiently large total number of collected particles $C_{s}$, the ratio $G_{s}/C_{s}$ is a constant, we can then calculate the time to collect $G_{s}$ particles. In the case where $u=0$, 


\begin{equation}\label{eq:TimeToCollectN}
t_{G}=\dfrac{G_{s}}{2\pi b^{2}n_{0}v_{T,s}},
\end{equation}
can be calculated for a given $G_{s}$ in the simulation, all other parameters being specified as initial conditions. The currents, $I_{s}$, of species $s=i,e$ are calculated by counting the amount of charge collected by the sphere, $eC_{s}$, and the time to collect that many particles $t_{G}$ using equation (\ref{eq:TimeToCollectN}),
\begin{equation}
\tilde{I_{s}}=\dfrac{I_{s}}{I_{s,0}}=\dfrac{eC_{s}}{t_{G}}\dfrac{1}{I_{s,0}}=\dfrac{C_{s}b_{i}^{2}}{2G_{s}}.
\end{equation}
Where the currents $\tilde{I_{s}}$ are normalised to the thermal current $I_{s,0}=4\pi a_{d}^{2}n_{0}e\sqrt{k_{B}T_{s}/2\pi m_{s}}$ which is the expected current of species $s$ through a spherical surface of radius $a_{d}$ in a field and flow free plasma. The measurement error is dependent simply on the number of particles recorded, $\sigma_{I_{s}}\propto\sqrt{\sigma_{N^{coll}_{s}}^{2}+\sigma_{N^{gen}_{s}}^{2}}$. In this mode, the dust grain acts effectively as a biased probe and can be benchmarked against previous numerical results from probe theory as discussed in the following section.


\subsection{Verification \& Validation}\label{Sec:VerificationAndValidation}
To give assurance to the implementation of the algorithm, the conservation of energy in simulations is verified. The ion and electron currents to the sphere in magnetised plasmas are measured and bench-marked against previous work. The results for the dependence of the normalised floating potential, $\chi=-e\phi_{d}/k_{B}T_{e}$, in an isothermal plasma in the weak and strong magnetic field limits are reproduced.

\begin{figure}
\centering
\begin{subfigure}[Example ion (red) and electron (blue) orbits close to a negatively charged, spherical dust grain.]{\includegraphics[width=0.5\textwidth]{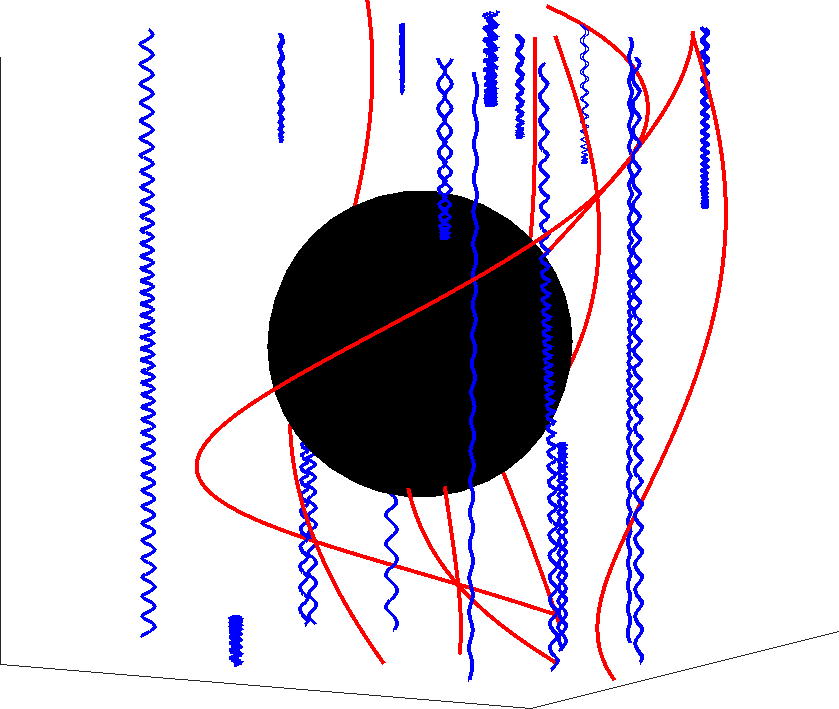}\label{Fig:DiMPlOrbits}}

\end{subfigure}
\begin{subfigure}[Example ion trajectories interacting weakly (red) and strongly (black) with dust grain with initial velocity $v_{z}=-0.2v_{T,i}$ impact parameters $b=1.5,2,2.5,3$ for $\beta_{i}=0.5$ and $\chi=2.5$ viewed perpendicular and parallel to $\underline{\widehat{B}}$.]{
\includegraphics[width=0.135\textwidth]{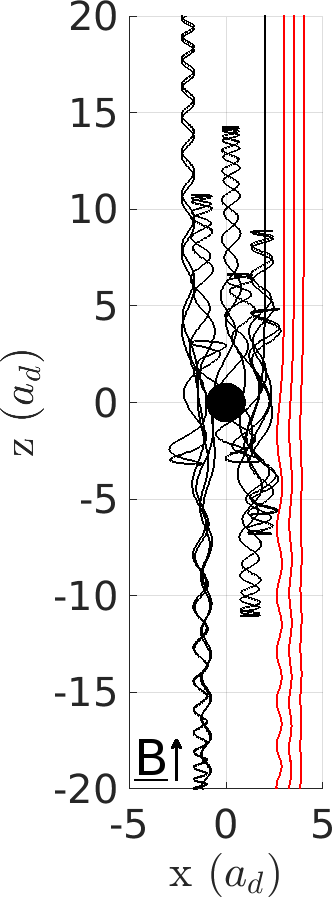}
\includegraphics[width=0.355\textwidth]{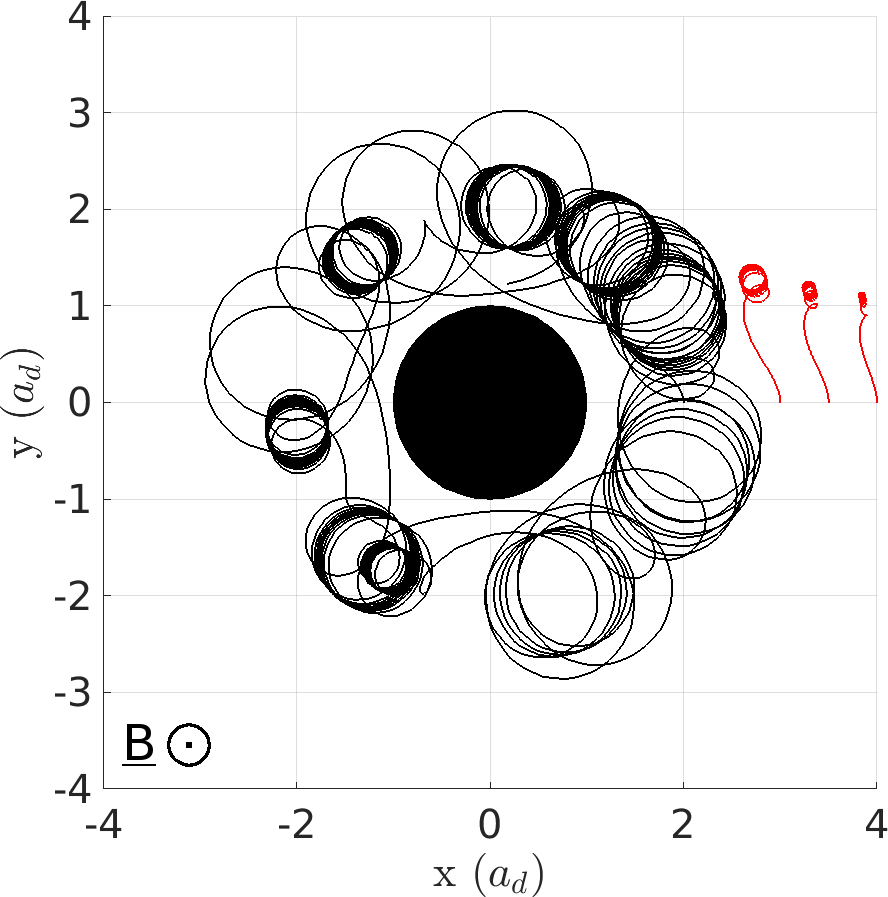}\label{Fig:TrappedOrbit}}
\end{subfigure}

\caption{Plots of particle orbits in DiMPl with a magnetic field in the vicinity of a negatively charged sphere with randomly directed velocity (a) and with velocity directed in the $\underline{\widehat{z}}$ direction from two orthogonal directions (b).}
\end{figure}

Example trajectories in DiMPl in figure \ref{Fig:DiMPlOrbits} show ion (red) and electron (blue) paths close to a negatively charged dust grain. The smaller inertia of electrons allows them to be significantly accelerated by the dust potential, and in many cases reflected. Figure \ref{Fig:TrappedOrbit} shows the trajectories of ions injected with a constant velocity $v_{z}=-0.2v_{T,s}$ across a range of impact parameters $b=1.5,2,2.5,3$ close to a negatively charged dust grain from two orthogonal directions, parallel and perpendicular to the magnetic field. The red lines show particles with large impact parameters which are displaced from a field line by the electrostatic attraction to the dust. For smaller impact parameters, ions can become transiently trapped, following complex paths as they reflect multiple times inside the attractive potential. In this way, ions exhibit an $\underline{E}\times\underline{B}$ drift due to the orthogonal electric and magnetic fields present at the $z=0$ plane.

The percentage deviation of the energy of ions in the absence of any electric fields is found to be of the order of machine precision $10^{-15}$. The change in energy when using the Boris algorithm is bounded, meaning local truncation errors dominate in this case. When an electric field is present, the variance in energy of an ion orbit is sensitively dependent on the exact particle trajectory however the greatest errors are found to be smaller than $\sim1\%$. 

\begin{figure}
\centering
\includegraphics[clip,width=0.5\textwidth]{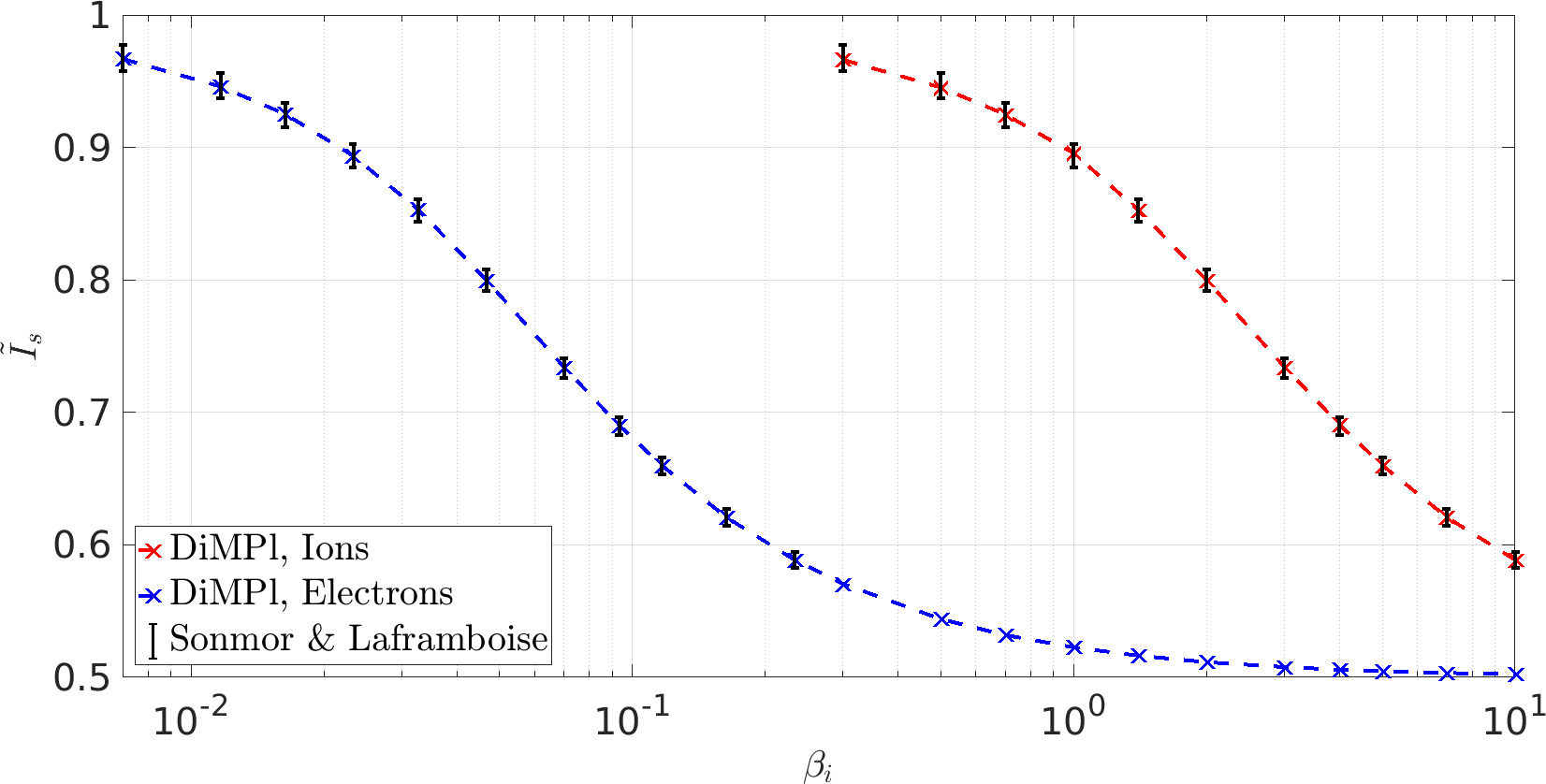}
\caption{Normalised ion and electron currents, $\tilde{I}_{s}$, as a function of the ion magnetisation parameter, $\beta_{i}$, as measured by DiMPl and compared with previous numerical results~\cite{Sonmor1991} for an uncharged dust grain with $z_{0}=1.01$ and $b=10.0$.}
\label{Fig:NoChargeCurrents}
\end{figure}

\begin{figure}
\centering
\includegraphics[clip,width=0.5\textwidth]{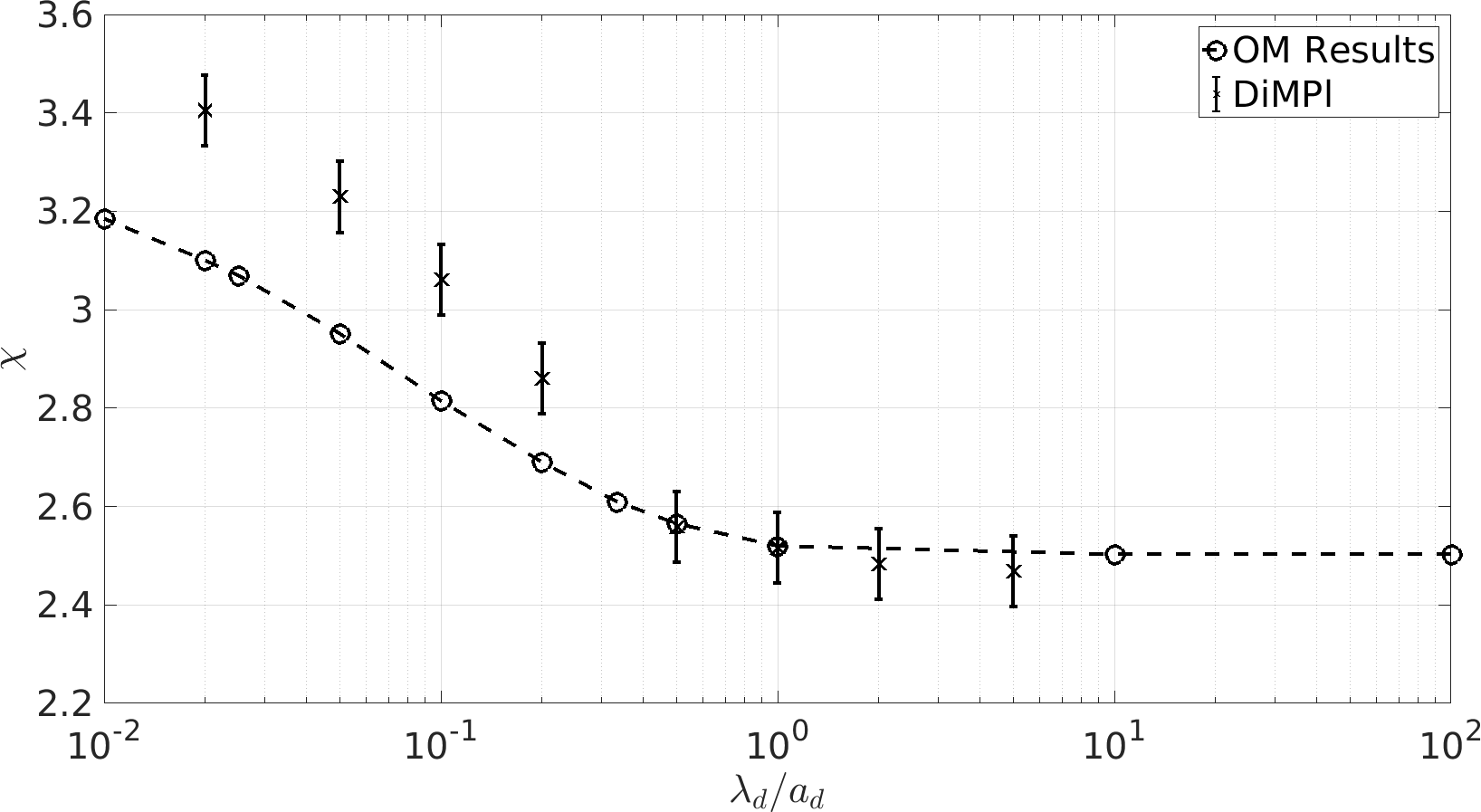}
\caption{Normalised potential, $\chi$, as a function of normalised Debye length, $\tilde{\lambda}_{d}$, as measured by DiMPl for ion magnetisation parameter $\beta_{i}=0.01$ and compared with numerical results of Kennedy~\cite{Kennedy2003} using $z_{0}=30$.}
\label{Fig:DiMPlSizeVerification}
\end{figure}

\begin{figure}
\centering
\includegraphics[clip,width=0.5\textwidth]{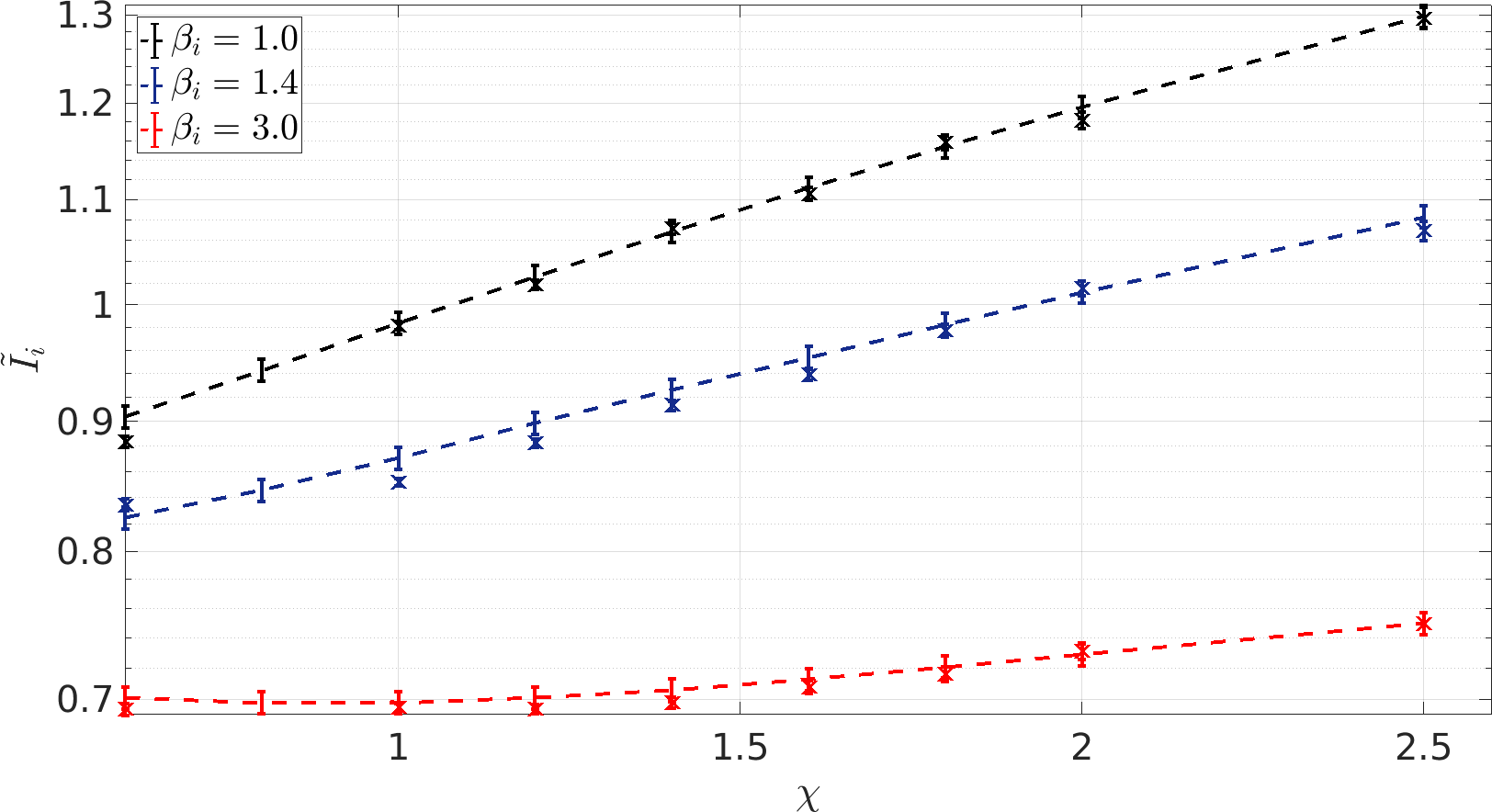}
\caption{The normalised ion current, $\tilde{I}_{i}$, as a function of the fixed potential bias, $\chi$, for $\beta_{i}=1.0,1.4,3.0$ as measured by DiMPl (markers) and compared with previous numerical results (dashed line)~\cite{Sonmor1991} using $b=3.0$.}
\label{Fig:AttChargedCurrents}
\end{figure}

Following the method described in section \ref{Sec:Method}, the currents of the attracted and repelled species as a function of probe potential and magnetic field strength were measured and compared to the previous numerical results of Sonmor and Laframboise~\cite{Sonmor1991}. The design of DiMPl makes it unsuitable for validation of the $B=0$ case. For an uncharged sphere, the results in figure \ref{Fig:NoChargeCurrents} shows excellent agreement for both electrons and ions over the full range of magnetic field strengths using $z_{0}=1.01$ and $b=10.0$. 

For large dust grain sizes comparable to the Debye length, the code was tested by comparison to previous numerical solutions~\cite{Kennedy2003} of orbital motion theory~\cite{Bohm1949,Bernstein1959,Laframboise1966,Allen1992}, as shown in figure \ref{Fig:DiMPlSizeVerification}. The accuracy of imposing an electric field following equations \ref{eq:DebyeFields} for a Debye-Huckel potential was evaluated for different normalised Debye lengths $\tilde{\lambda}_{d}$ with DiMPl in the low magnetic field limit, $\beta_{i}=0.01$. The parameter $\tilde{\lambda}_{d}$ was varied in DiMPl by altering the dust grain size whilst keeping all other parameters fixed with $z_{0}=30$. The floating potential calculated using DiMPl for different values of $\tilde{\lambda}_{d}$ shows good agreement for $\tilde{\lambda}_{d}\geq0.5$, supporting the approximation of a Debye-Huckel potential following equation \ref{eq:DebyeFields} in this regime.

Figure \ref{Fig:AttChargedCurrents} shows the variation of the normalised attracted species current for different floating potentials, for three values of the ion magnetisation parameter $\beta_{i}=1.0,1.4,3.0$. These results show good agreement with previous numerical results for the current~\cite{Sonmor1991}. This justifies the use of equation \ref{eq:TimeToCollectN} for calculating the current for a fixed surface potential. Results for the floating potentials and currents were found to agree well with previous results~\cite{Patacchini2007,Sonmor1991} over the range $0.3\leq\beta_{i}\leq10.0$.

\begin{figure}
\centering
\includegraphics[clip,width=0.5\textwidth]{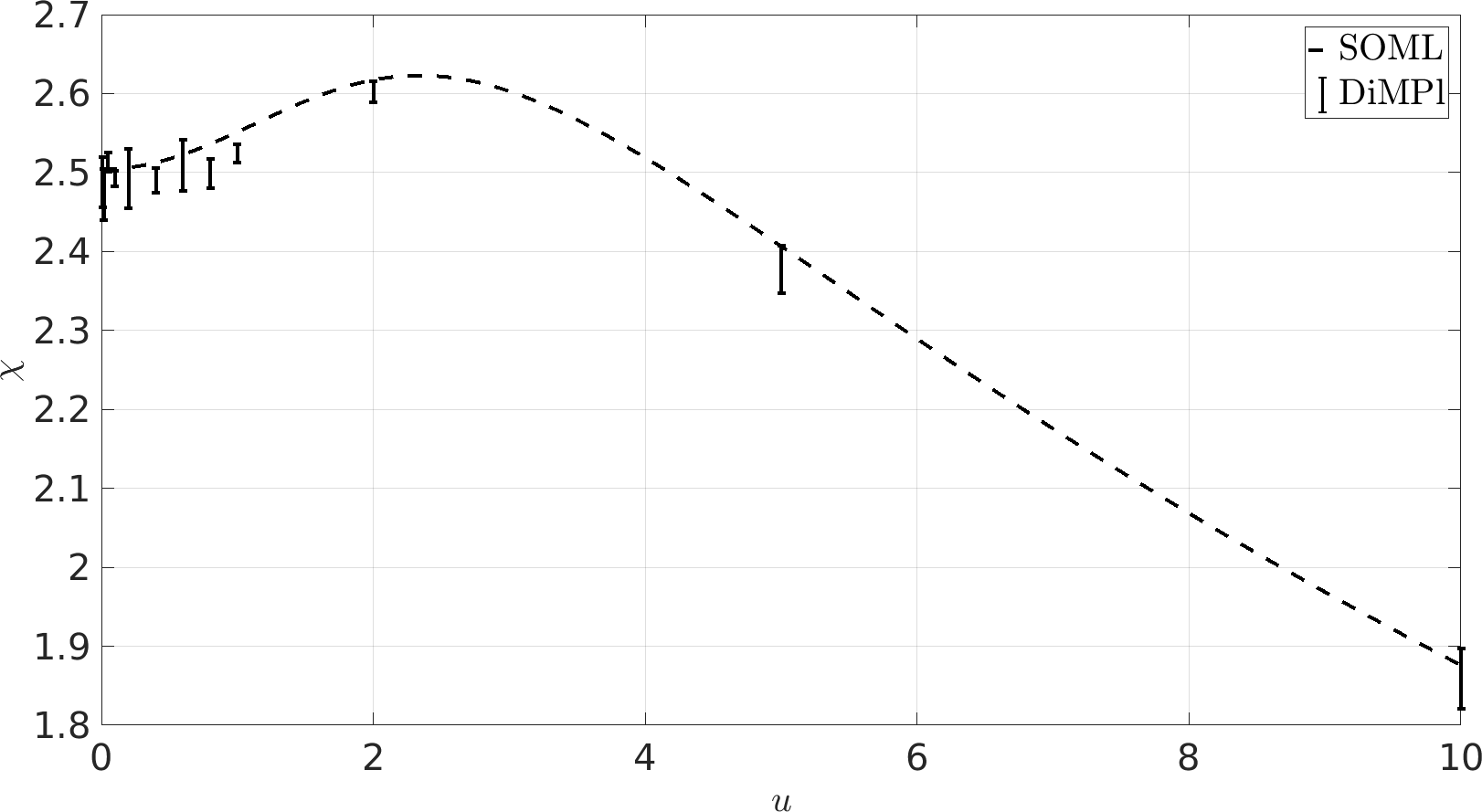}
\caption{The normalised floating potential, $\chi$, as a function of the normalised flow velocity, $u$, for $\beta_{i}=0.01$ and $z_{0}=100.0$ as measured by DiMPl and compared with predictions of SOML theory.}
\label{Fig:DiMPlSOMLValidation_Cropped}
\end{figure}

The results of DiMPl for a coulomb potential are expected to conform with the predictions of SOML theory since they utilise the same assumptions. Figure \ref{Fig:DiMPlSOMLValidation_Cropped} shows the normalised potential as a function of flow speed of DiMPl as compared with SOML theory. The results presented provide evidence which support the application of the simulation methodology.


\section{\label{sec:Simulation results}Simulation results}
Here, the results of DiMPl are compared with measurements made with the Plasma Oct Tree code \texttt{pot}~\cite{Thomas2016} and the previous PIC code results of Lange~\cite{Lange2016}. Similar to DiMPl, \texttt{pot} solves an N-body simulation of electrons and ions in proximity to a charged, conducting spheroid. The advantage of \texttt{pot} is that it solves the entire N-body problem, making the result inherently self-consistent in it's implementation of collisional and sheath effects. The principle assumption is that electrons and ions arrive from the distant plasma uniformly distributed over a spherical surface with a Maxwellian velocity distribution. Each individual particle is tracked in a constant uniform magnetic field and an electric field sensitive to the influence of all other particles in the simulation as well as the central charged sphere.

\begin{figure}[h]
\centering
\includegraphics[clip,width=0.5\textwidth]{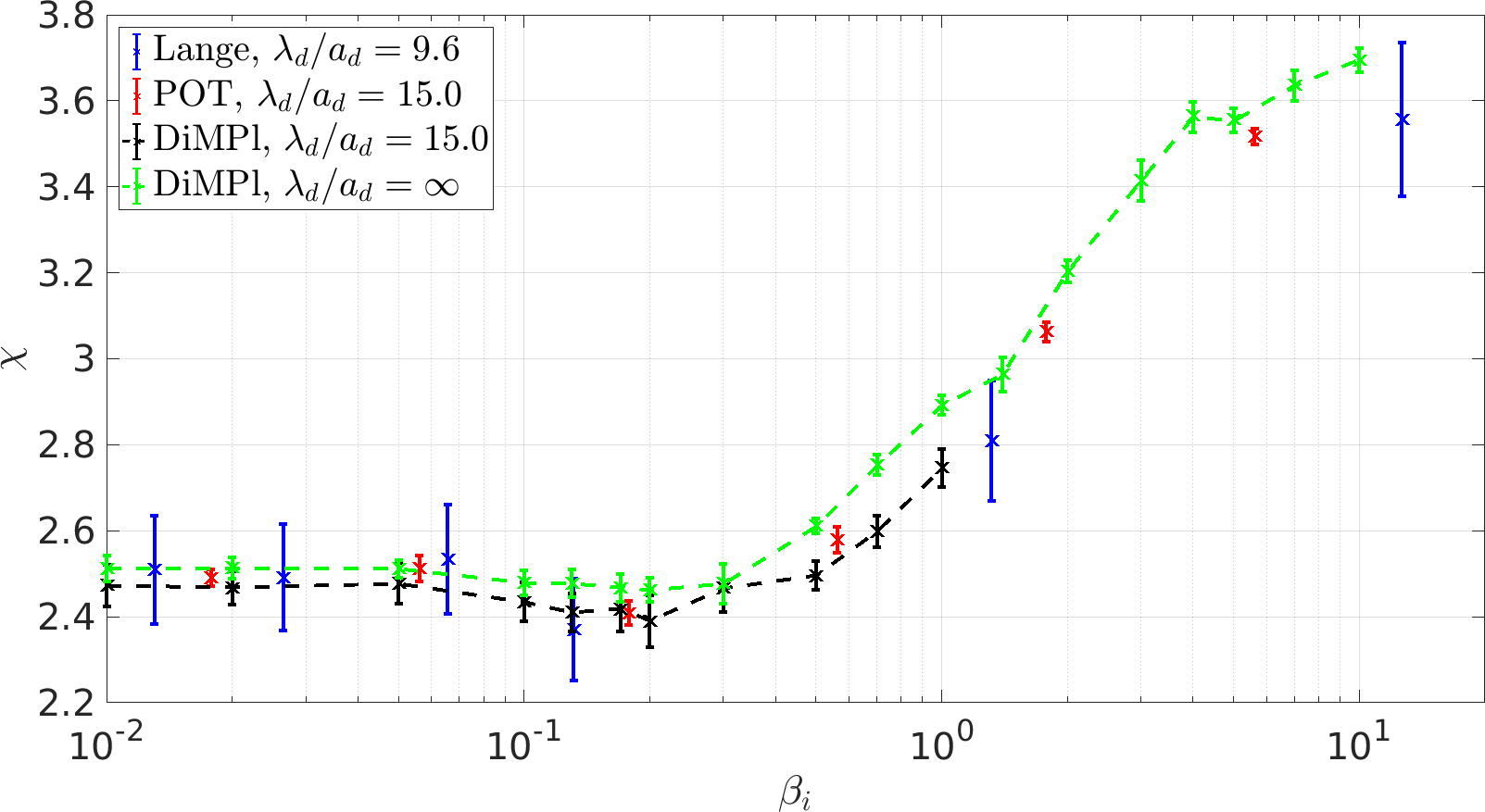}
\caption{Normalised potential, $\chi$, as a function of the ion magnetisation parameter, $\beta_{i}$, as measured by DiMPl for a Debye-huckel potential (black) using equation (\ref{eq:DebyeFields}) and \texttt{pot}~\cite{Thomas2016} for a normalised Debye length of $\tilde{\lambda}_{d}=15$ (red), alongside the results of Lange~\cite{Lange2016} for $\tilde{\lambda}_{d}=9.6$ (blue) and DiMPl using equation (\ref{eq:CoulombFields}) (green).}
\label{Fig:DiMPlMagnetisationVerification}
\end{figure}

In figure \ref{Fig:DiMPlMagnetisationVerification}, measurements of the floating potential made with DiMPl using a Debye-Huckel electric field from equation \ref{eq:DebyeFields} with a normalised Debye length $\tilde{\lambda}_{d}=15$ and for a coulomb potential using equation \ref{eq:CoulombFields} were compared against the results of \texttt{pot} for $\tilde{\lambda}_{d}=15$ and the PIC code results of Lange~\cite{Lange2016} for the closest value of $\tilde{\lambda}_{d}=9.598$. In \texttt{pot}, this value of $\tilde{\lambda}_{d}$ was attained by changing the plasma density whilst maintaining $n_{i}=n_{e}$ and $T_{i}=T_{e}$. The results of Lange have been plotted with a $5\%$ error to reflect the reported systematic error and correction made in plotting here~\cite{Lange2016}. 

\begin{figure}
\centering
\includegraphics[trim={0cm 0cm 0cm 0cm},clip,width=0.5\textwidth]{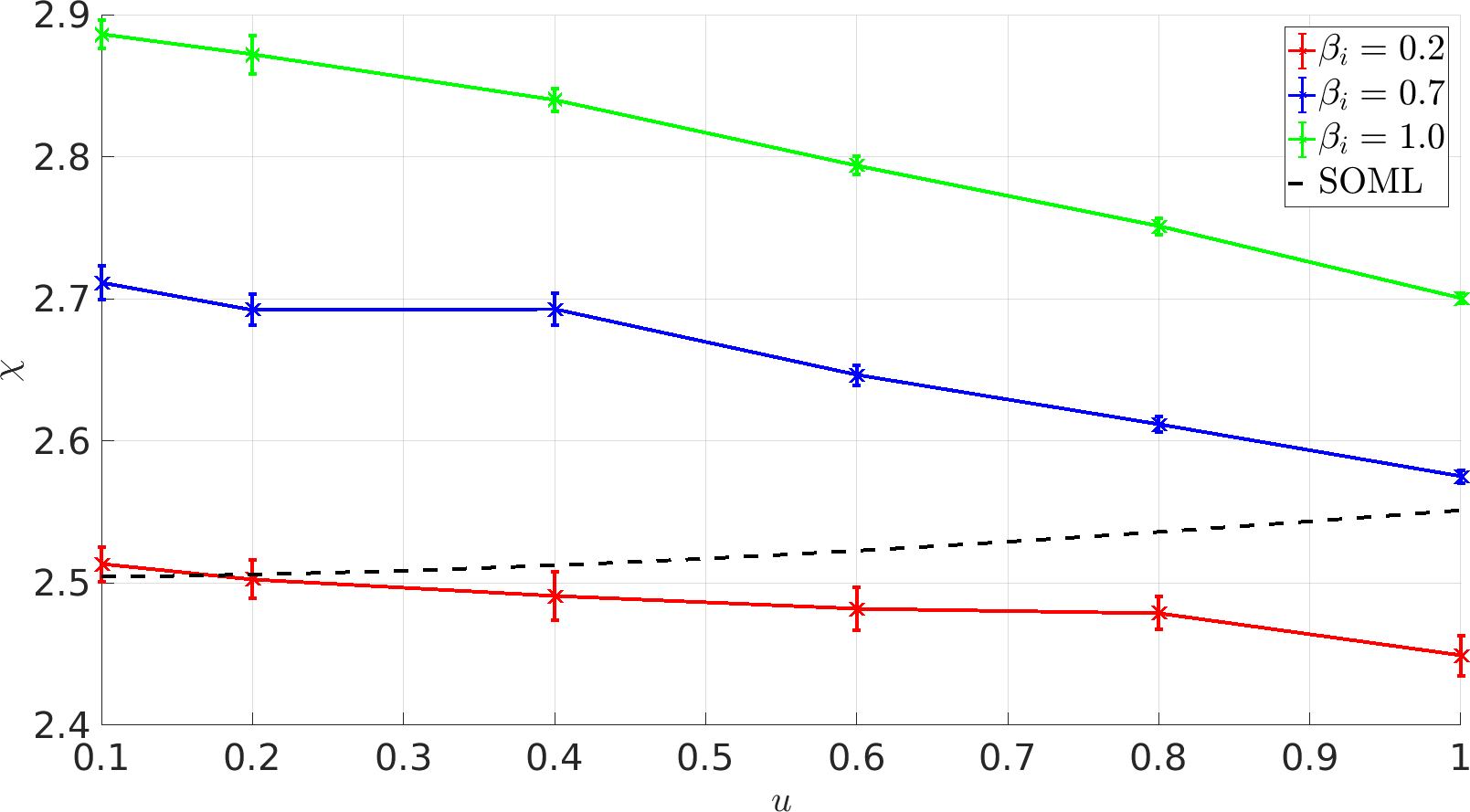}
\caption{Normalised potential, $\chi$, as a function of flow velocity as measured by DiMPl (solid line) for $\beta_{i}=0.01,0.2,0.7,1.0$ as compared with the results of SOML theory (dashed line) with $b=3.0$.}
\label{Fig:DiMPlCoulombFlow_Cropped}
\end{figure}

\begin{figure}

\begin{subfigure}[$\tilde{\lambda}_{d}=1.0$.]{\includegraphics[trim={0cm 0cm 0cm 0cm},clip,width=0.5\textwidth]{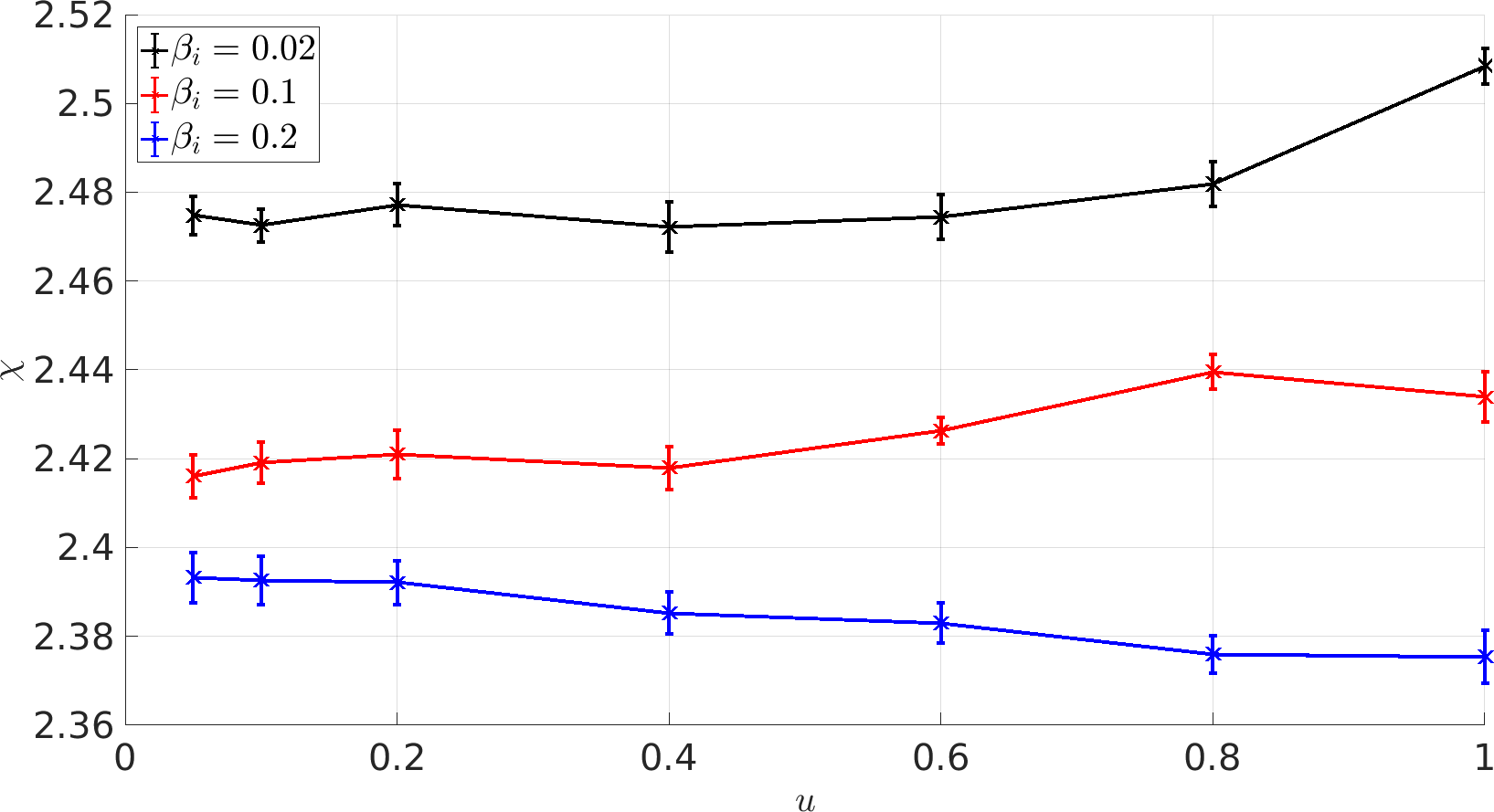}\label{Fig:DiMPlDebyeFlowr7mu_Cropped}}

\end{subfigure}

\begin{subfigure}[$\tilde{\lambda}_{d}=7.4$.]{\includegraphics[trim={0cm 0cm 0cm 0cm},clip,width=0.5\textwidth]{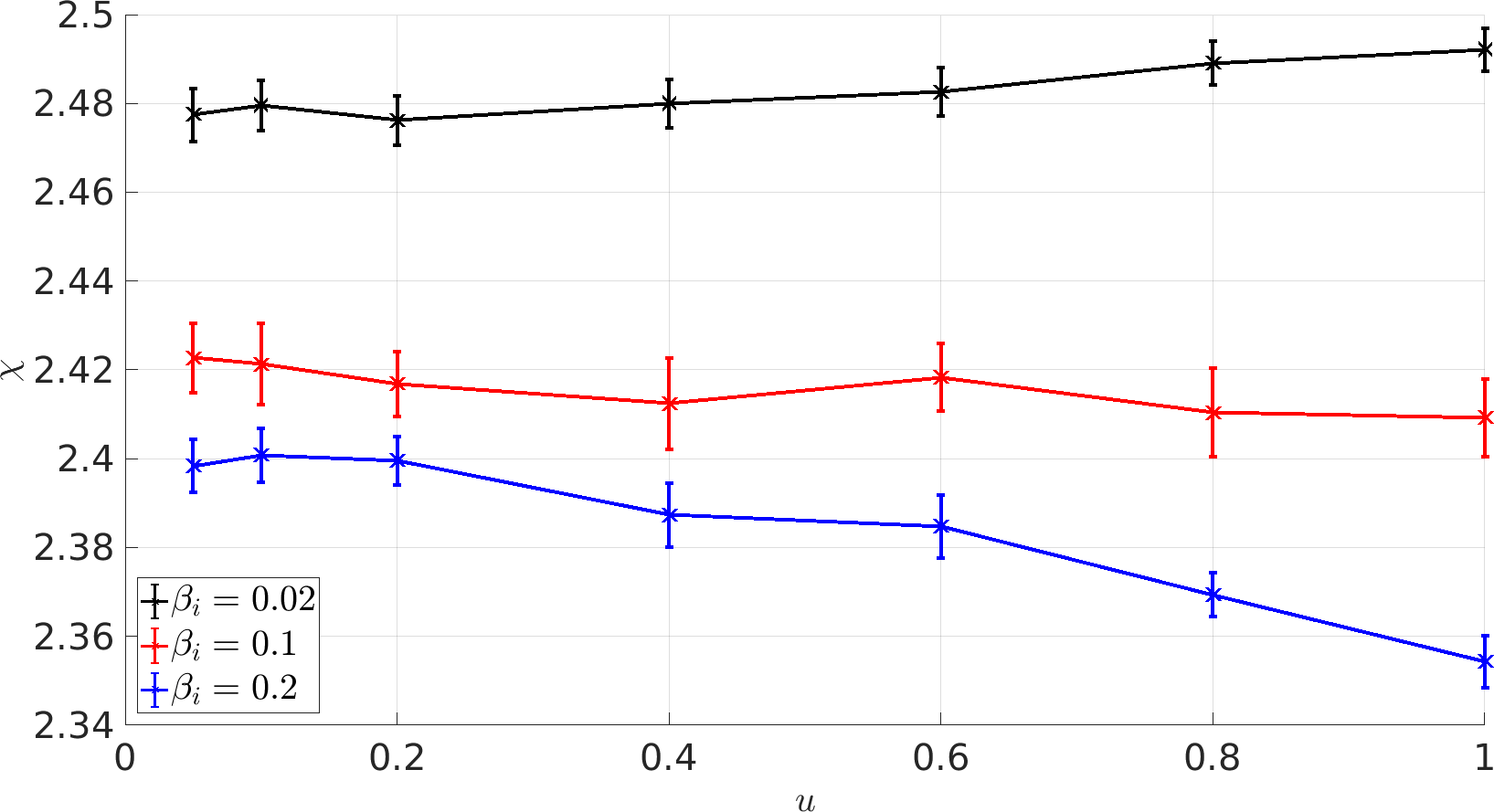}\label{Fig:DiMPlDebyeFlowr1mu_Cropped}}
\centering

\end{subfigure}
\caption{Normalised potential, $\chi$, as a function of flow velocity as measured by DiMPl for $\beta_{i}=0.02,0.1,0.2$ for large dust $\tilde{\lambda}_{d}=1.0,7.4$ for electric fields calculated using equation (\ref{eq:DebyeFields}) with $z_{0}=30.0$ and $b=3.0$.}
\end{figure}

All three codes predict the same qualitative trend for $\tilde{\lambda}_{d}\gg 1$, with a slight reduction in potential around $\beta_{i}\simeq0.15$ followed by a gradual increase to much higher potentials for $\beta_{i}>1$. All three codes reproduce the expected OML result at low magnetisation $\beta_{i}=0.01$ of $\chi=2.50$. The uncertainty in the simulation results of Lange are large enough to account for the trends seen by all other results. The data of DiMPl and \texttt{pot} are in agreement for $\lambda_{d}/a_{d}=15$ over the examined range. When using a coulomb potential in DiMPl the results produced are very similar to the case of $\tilde{\lambda}_{d}=15$, however in the high field limit the potential is slightly elevated.

DiMPl has also been employed to investigate the effect of magnetic fields and flow on the floating potential of dust grains of a range of sizes. Using a coulomb potential with $b=3.0$ the floating potential of small dust grains, $\tilde{\lambda}_{d}\gg1$, was measured for a range of flow speeds between $u=0.1$ and $u=1.0$ and ion magnetisation parameters between $\beta_{i}=0.01$ and $\beta_{i}=1.0$. The results shown in figure \ref{Fig:DiMPlCoulombFlow_Cropped}, found that the presence of a magnetic field causes the potential to decrease with increasing flow speed. This effect is enhanced with increasing magnetisation as the potential is observed to decrease more rapidly.

Results were gathered for larger dust sizes, $\tilde{\lambda}_{d}=7.4$ and $\tilde{\lambda}_{d}=1.0$, for ion magnetisation parameters between $\beta_{i}=0.01$ and $\beta_{i}=0.2$ using $z_{0}=30.0$ and $b=3.0$. The results show that the potential is slightly decreased and depends less strongly than small dust on flow velocity, as shown in figure \ref{Fig:DiMPlDebyeFlowr7mu_Cropped} and \ref{Fig:DiMPlDebyeFlowr1mu_Cropped}. The surface potential of large dust is found to be approximately independent of flow speed for low magnetic field strengths.

\section{\label{sec:Semi-empirical formulation of floating potential}Semi-empirical formulation of floating potential}
Guided by the theoretical understanding of the underlying effect of magnetic fields on the flux of ions and electrons to the dust~\cite{Tsytovich2003}, a semi-empirical formulation for the potential as a function of ion magnetisation is presented. In an isothermal hydrogen plasma, these results and others indicate that initially the magnitude of the potential becomes more positive with increasing magnetic field strength as the electron gyro-radius becomes comparable with the sphere radius, before then approaching the high field limit as the ions also become strongly magnetised~\cite{Patacchini2007,Lange2016,Thomas2016}. More formally, in the limit of $a_{d}\ll\lambda_{d}$ and $a_{d}\ll \rho_{\perp,e},\rho_{\perp,i}$, we expect to recover the well known OML result for the ion and electron current to a negatively charged sphere
\begin{equation}
\begin{split}
&I_{i}(\beta\rightarrow0)=I_{i,0}\Big(1+\dfrac{Z}{\tau}\chi\Big), \\
&I_{e}(\beta\rightarrow0)=I_{e,0}e^{-\chi}.
\end{split}
\label{eq:CurrentBzeroLinf}
\end{equation}
with mean ionisation $Z$ and temperature ratio $\tau=T_{i}/T_{e}$. In the opposite magnetic field limit, $a_{d}\ll\lambda_{d}$ and $a_{d}\gg \rho_{\perp,e},\rho_{\perp,i}$, we expect all plasma species to follow trajectories along magnetic field lines such that only ions incident over the projected area of the sphere along the $z$ axis are collected. For electrons a portion of them are reflected following a Boltzmann distribution, meaning
\begin{equation}
\begin{split}
&I_{i}(\beta\rightarrow\infty)=\dfrac{1}{2}I_{i,0}, \\
&I_{e}(\beta\rightarrow\infty)=\dfrac{1}{2}I_{e,0}e^{-\chi}.
\end{split}
\label{eq:CurrentBinfLinf}
\end{equation}
Combining equations (\ref{eq:CurrentBzeroLinf}) and (\ref{eq:CurrentBinfLinf}) with parameterised exponential functions,
\begin{equation}\label{eq:MagnetisedCurrents}
\begin{split}
&I_{i}=\dfrac{I_{i,0}}{2}\Big[\Big(1+\dfrac{2Z\chi}{\tau}\Big)e^{-\alpha_{2}\beta_{i}^{\alpha_{3}}}+1\Big], \\
&I_{e}=\dfrac{I_{e,0}}{2}e^{-\chi}\Big[e^{-\alpha_{1}\beta_{e}}+1\Big],
\end{split}
\end{equation}
this formulation has the desired properties of approaching the current limits $I_{s}(\beta_{i}\rightarrow0)$ and $I_{s}(\beta_{i}\rightarrow\infty)$, where $\alpha_{1}$, $\alpha_{2}$ and $\alpha_{3}$ are free parameters and $\beta_{e}=\sqrt{\mu\tau}\beta_{i}$. The electron and ion currents of equation (\ref{eq:MagnetisedCurrents}) can be fit well to the numerical results of Sonmor \& Laframboise~\cite{Sonmor1991} over the range $0.3\leq\beta_{i}\leq10.0$ and $0\leq\chi\leq5.0$.

\begin{figure}[h]
\centering
\includegraphics[clip,width=0.5\textwidth]{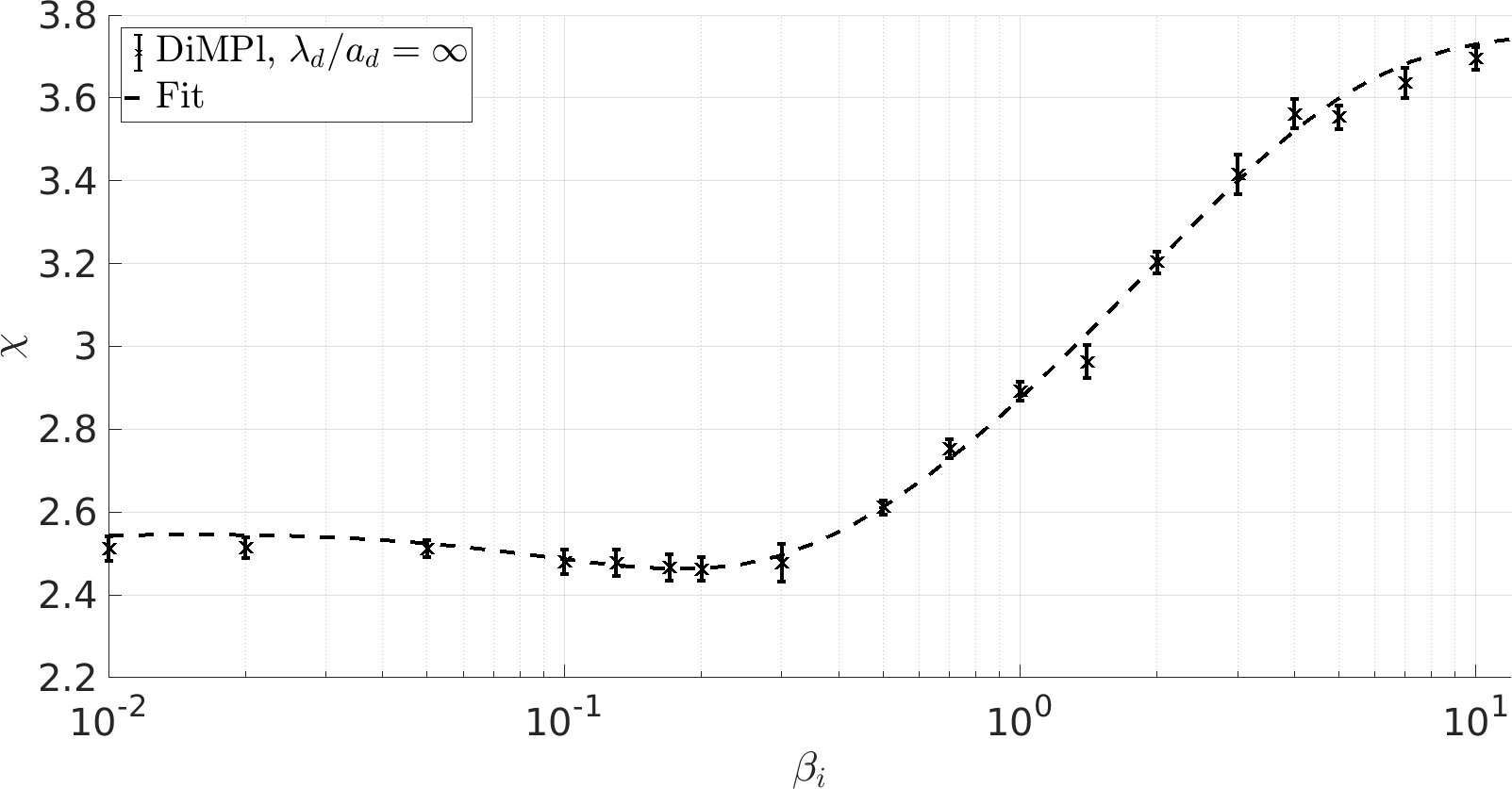}
\caption{Fit to the normalised surface potential, $\chi$, (dashed) as a function of the ion magnetisation parameter, $\beta_{i}$, following equation (\ref{eq:MagnetisedCurrents}) as compared with DiMPl with a Coulomb potential. }
\label{Fig:Semi-EmpiricalFit_DiMPl}
\end{figure}

\begin{figure}[h]
\centering
\includegraphics[clip,width=0.5\textwidth]{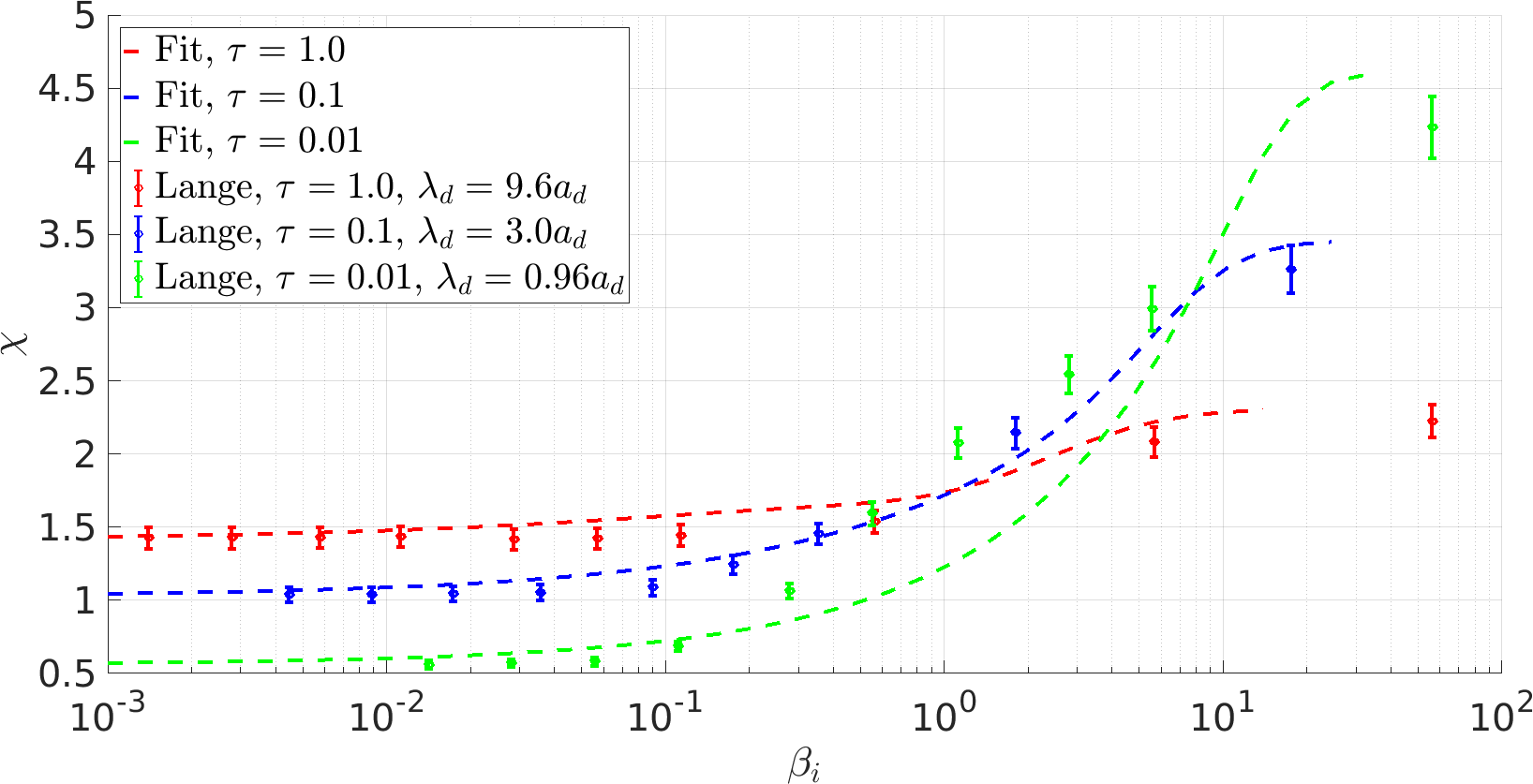}
\caption{Results for the normalised surface potential, $\chi$, recorded by Lange~\cite{Lange2016} (markers) fit using equation \ref{eq:MagnetisedCurrents} (dashed) for mass ratio $\mu=100$ and temperature ratios $\tau=1.0$, $0.1$ and $0.01$. The fit recovers the trend for $\tau=0.1$ but inaccurately estimates the potential for $\tau=0.01$. }
\label{Fig:Semi-EmpiricalFit_Lange}
\end{figure}

For the values $\alpha_{1}=0.23$, $\alpha_{2}=1.56$ and $\alpha_{3}=0.56$, equation \ref{eq:MagnetisedCurrents} fits the DiMPl results with a coulomb potential extremely well with an $R^{2}=0.996$ and RMSE of $0.006$ as shown in figure \ref{Fig:Semi-EmpiricalFit_DiMPl}. For $\tilde{\lambda}_{d}=\infty$, this produces an accurate description of the surface potential dependence on magnetic field strength parameterised through $\beta_{i}$ which, as expected, recovers the expected potential in the limits $\beta_{i}\rightarrow0$ and $\beta_{i}\rightarrow\infty$. 

The semi-empirical model has been applied in predicting the results of Lange for a mass ratio $\mu=100$ and for three values of temperature ratio $\tau=1.0$, $\tau=0.1$ and $\tau=0.01$, as shown in figure \ref{Fig:Semi-EmpiricalFit_Lange}. The model accurately predicts the results for $\tau=1.0$ and the trend for $\tau=0.1$ though there is a discrepancy in a small range around $\beta_{i}=1.0$. For $\tau=0.01$ the model breaks down and misses the onset of magnetic effects at $\beta_{i}\geq0.1$.

\section{Conclusions}
Understanding the charging behaviour of conducting dust grains in magnetised collisionless plasmas is vital to predicting their dynamics and consequently for interpreting astrophysical measurements and controlling impurity deposition in tokamaks. The Monte Carlo code DiMPl, designed to simulate the forces and charge accumulation experienced by a dust grain in a collisionless magnetised plasma has been presented. The results have been validated against the full N-body simulations of \texttt{pot}~\cite{Thomas2016}, PIC code results~\cite{Lange2016} and other previous numerical results~\cite{Sonmor1991}. In particular, the dependency of the ion and electron currents and the floating surface potential $\chi$ on the ion magnetisation parameter $\beta_{i}$ for small dust grains with $\tilde{\lambda}\gg1$ have been found to agree with previous numerical solutions. In comparison to predictions from OM calculations, DiMPl was found to deviate in an isothermal plasma only for $\tilde{\lambda}_{d}<0.5$. The measurements of DiMPl for the dependence of the floating potential on magnetic field strength are in agreement with other PIC and Monte Carlo code simulations~\cite{Lange2016,Patacchini2007,Thomas2016}. The results of DiMPl suggest a small region where the electron current is more significantly influenced by the magnetic field than the ion current at a magnetisation parameter of $\beta_{i}\simeq0.15$, causing a dip in the surface potential before a gradual increase with magnetic field as the ions become significantly affected. The equilibrium surface potential with a Debye-Huckel potential for $\tilde{\lambda}=15.0$ was found to follow a similar trend to the results for a coulomb potential, with a small decrease in potential at high magnetic field strengths bringing it into better agreement with the results of \texttt{pot} for the same dust size. 

For flowing plasmas, the results of DiMPl are in agreement with SOML theory in the low field limit. For increasing magnetic field strength, the potential decreases in magnitude up to $\beta_{i}=0.2$ and then increases for $\beta_{i}>0.2$, which is most likely due to the decrease in ion current. For larger dust with $\tilde{\lambda}_{d}=7.4,1.0$, the presence of a magnetic field was found to reduce the dependence on flow and the potential overall was decreased with increasing magnetic field strength up to $\beta_{i}=0.2$.

A semi-empirical model for the dependency of the floating potential on the ion magnetisation parameter $\beta_{i}$ has been presented for a collisionless, fully ionised, hydrogen plasma. The model performs well as compared with available simulation results for $\tau>0.01$ and recovers the correct theoretical model in the limit of high and low $\beta_{i}$.

The omission of collisional effects and approximation of spherically symmetric potentials in DiMPl permits far shorter computation timescales in comparison to \texttt{pot} and PIC code simulations. The potential formulation in DiMPl does not provide a solution which is self-consistent with the plasma behaviour around a charged conductor. However, the results presented show surprisingly good agreement with the expected dependence of the surface potential on magnetic field strength for a range of dust sizes, suggesting that these assumptions provide accurate approximations in the regime $\tilde{\lambda}_{d}\geq0.5$.

This work has been supported by the UK's Engineering and Physical Sciences Research Council. The data that support the findings of this study are available from the corresponding author upon reasonable request.

\bibliography{FloatingPotentialOfSphericalDustInCollisionlessMagnetisedPlasmas}

\begin{thebibliography}{47}%
\makeatletter
\providecommand \@ifxundefined [1]{%
 \@ifx{#1\undefined}
}%
\providecommand \@ifnum [1]{%
 \ifnum #1\expandafter \@firstoftwo
 \else \expandafter \@secondoftwo
 \fi
}%
\providecommand \@ifx [1]{%
 \ifx #1\expandafter \@firstoftwo
 \else \expandafter \@secondoftwo
 \fi
}%
\providecommand \natexlab [1]{#1}%
\providecommand \enquote  [1]{``#1''}%
\providecommand \bibnamefont  [1]{#1}%
\providecommand \bibfnamefont [1]{#1}%
\providecommand \citenamefont [1]{#1}%
\providecommand \href@noop [0]{\@secondoftwo}%
\providecommand \href [0]{\begingroup \@sanitize@url \@href}%
\providecommand \@href[1]{\@@startlink{#1}\@@href}%
\providecommand \@@href[1]{\endgroup#1\@@endlink}%
\providecommand \@sanitize@url [0]{\catcode `\\12\catcode `\$12\catcode
  `\&12\catcode `\#12\catcode `\^12\catcode `\_12\catcode `\%12\relax}%
\providecommand \@@startlink[1]{}%
\providecommand \@@endlink[0]{}%
\providecommand \url  [0]{\begingroup\@sanitize@url \@url }%
\providecommand \@url [1]{\endgroup\@href {#1}{\urlprefix }}%
\providecommand \urlprefix  [0]{URL }%
\providecommand \Eprint [0]{\href }%
\providecommand \doibase [0]{http://dx.doi.org/}%
\providecommand \selectlanguage [0]{\@gobble}%
\providecommand \bibinfo  [0]{\@secondoftwo}%
\providecommand \bibfield  [0]{\@secondoftwo}%
\providecommand \translation [1]{[#1]}%
\providecommand \BibitemOpen [0]{}%
\providecommand \bibitemStop [0]{}%
\providecommand \bibitemNoStop [0]{.\EOS\space}%
\providecommand \EOS [0]{\spacefactor3000\relax}%
\providecommand \BibitemShut  [1]{\csname bibitem#1\endcsname}%
\let\auto@bib@innerbib\@empty
\bibitem [{\citenamefont {Mott-Smith}\ and\ \citenamefont
  {Langmuir}(1926)}]{Mott-Smith1926}%
  \BibitemOpen
  \bibfield  {author} {\bibinfo {author} {\bibfnamefont {H.~M.}\ \bibnamefont
  {Mott-Smith}}\ and\ \bibinfo {author} {\bibfnamefont {I.}~\bibnamefont
  {Langmuir}},\ }\bibfield  {title} {\enquote {\bibinfo {title} {{The theory of
  collectors in gaseous discharges}},}\ }\href {\doibase
  10.1103/PhysRev.28.727} {\bibfield  {journal} {\bibinfo  {journal} {Physical
  Review}\ }\textbf {\bibinfo {volume} {28}},\ \bibinfo {pages} {727--763}
  (\bibinfo {year} {1926})}\BibitemShut {NoStop}%
\bibitem [{\citenamefont {Mendis}\ and\ \citenamefont
  {Rosenberg}(1994)}]{Mendis1994}%
  \BibitemOpen
  \bibfield  {author} {\bibinfo {author} {\bibfnamefont {D.~A.}\ \bibnamefont
  {Mendis}}\ and\ \bibinfo {author} {\bibfnamefont {M.}~\bibnamefont
  {Rosenberg}},\ }\bibfield  {title} {\enquote {\bibinfo {title} {{COSMIC DUSTY
  PLASMA}},}\ }\href@noop {} {\bibfield  {journal} {\bibinfo  {journal} {Annu.
  Rev. Astron. Astrophys}\ }\textbf {\bibinfo {volume} {32}},\ \bibinfo {pages}
  {419--463} (\bibinfo {year} {1994})}\BibitemShut {NoStop}%
\bibitem [{\citenamefont {Hopkins}\ and\ \citenamefont
  {Graham}(1986)}]{Hopkins1986}%
  \BibitemOpen
  \bibfield  {author} {\bibinfo {author} {\bibfnamefont {M.~B.}\ \bibnamefont
  {Hopkins}}\ and\ \bibinfo {author} {\bibfnamefont {W.~G.}\ \bibnamefont
  {Graham}},\ }\bibfield  {title} {\enquote {\bibinfo {title} {{Langmuir probe
  technique for plasma parameter measurement in a medium density discharge}},}\
  }\href {\doibase 10.1063/1.1138684} {\bibfield  {journal} {\bibinfo
  {journal} {Review of Scientific Instruments}\ }\textbf {\bibinfo {volume}
  {57}},\ \bibinfo {pages} {2210--2217} (\bibinfo {year} {1986})}\BibitemShut
  {NoStop}%
\bibitem [{\citenamefont {Spitzer}(1941)}]{Spitzer1941}%
  \BibitemOpen
  \bibfield  {author} {\bibinfo {author} {\bibfnamefont {L.}~\bibnamefont
  {Spitzer}},\ }\bibfield  {title} {\enquote {\bibinfo {title} {{The dynamics
  of the interstellar medium}},}\ }\href@noop {} {\bibfield  {journal}
  {\bibinfo  {journal} {The Astrophysical Journal}\ }\textbf {\bibinfo {volume}
  {93}},\ \bibinfo {pages} {369--379} (\bibinfo {year} {1941})}\BibitemShut
  {NoStop}%
\bibitem [{\citenamefont {Bacharis}, \citenamefont {Coppins},\ and\
  \citenamefont {Allen}(2010)}]{Bacharis2010}%
  \BibitemOpen
  \bibfield  {author} {\bibinfo {author} {\bibfnamefont {M.}~\bibnamefont
  {Bacharis}}, \bibinfo {author} {\bibfnamefont {M.}~\bibnamefont {Coppins}}, \
  and\ \bibinfo {author} {\bibfnamefont {J.~E.}\ \bibnamefont {Allen}},\
  }\bibfield  {title} {\enquote {\bibinfo {title} {{Critical issues for
  modeling dust transport in tokamaks}},}\ }\href {\doibase
  10.1103/PhysRevE.82.026403} {\bibfield  {journal} {\bibinfo  {journal}
  {Physical Review E - Statistical, Nonlinear, and Soft Matter Physics}\
  }\textbf {\bibinfo {volume} {82}},\ \bibinfo {pages} {1--5} (\bibinfo {year}
  {2010})}\BibitemShut {NoStop}%
\bibitem [{\citenamefont {Garrett}(1981)}]{Garrett1981}%
  \BibitemOpen
  \bibfield  {author} {\bibinfo {author} {\bibfnamefont {H.~B.}\ \bibnamefont
  {Garrett}},\ }\bibfield  {title} {\enquote {\bibinfo {title} {{The charging
  of spacecraft surfaces}},}\ }\href {\doibase 10.1029/RG019i004p00577}
  {\bibfield  {journal} {\bibinfo  {journal} {Reviews of Geophysics}\ }\textbf
  {\bibinfo {volume} {19}},\ \bibinfo {pages} {577--616} (\bibinfo {year}
  {1981})}\BibitemShut {NoStop}%
\bibitem [{\citenamefont {Olson}\ \emph {et~al.}(2010)\citenamefont {Olson},
  \citenamefont {Miloch}, \citenamefont {Ratynskaia},\ and\ \citenamefont
  {Yaroshenko}}]{Olson2010}%
  \BibitemOpen
  \bibfield  {author} {\bibinfo {author} {\bibfnamefont {J.}~\bibnamefont
  {Olson}}, \bibinfo {author} {\bibfnamefont {W.~J.}\ \bibnamefont {Miloch}},
  \bibinfo {author} {\bibfnamefont {S.}~\bibnamefont {Ratynskaia}}, \ and\
  \bibinfo {author} {\bibfnamefont {V.}~\bibnamefont {Yaroshenko}},\ }\bibfield
   {title} {\enquote {\bibinfo {title} {{Potential structure around the Cassini
  spacecraft near the orbit of Enceladus}},}\ }\href {\doibase
  10.1063/1.3486523} {\bibfield  {journal} {\bibinfo  {journal} {Physics of
  Plasmas}\ }\textbf {\bibinfo {volume} {17}} (\bibinfo {year} {2010}),\
  10.1063/1.3486523}\BibitemShut {NoStop}%
\bibitem [{\citenamefont {Stangeby}(1984)}]{Stangeby1984}%
  \BibitemOpen
  \bibfield  {author} {\bibinfo {author} {\bibfnamefont {P.~C.}\ \bibnamefont
  {Stangeby}},\ }\bibfield  {title} {\enquote {\bibinfo {title} {{Plasma sheath
  transmission factors for tokamak edge plasmas}},}\ }\href {\doibase
  10.1063/1.864677} {\bibfield  {journal} {\bibinfo  {journal} {Physics of
  Fluids}\ }\textbf {\bibinfo {volume} {27}},\ \bibinfo {pages} {682} (\bibinfo
  {year} {1984})}\BibitemShut {NoStop}%
\bibitem [{\citenamefont {Malizia}\ \emph {et~al.}(2016)\citenamefont
  {Malizia}, \citenamefont {Poggi}, \citenamefont {Ciparisse}, \citenamefont
  {Rossi}, \citenamefont {Bellecci},\ and\ \citenamefont
  {Gaudio}}]{Malizia2016}%
  \BibitemOpen
  \bibfield  {author} {\bibinfo {author} {\bibfnamefont {A.}~\bibnamefont
  {Malizia}}, \bibinfo {author} {\bibfnamefont {L.~A.}\ \bibnamefont {Poggi}},
  \bibinfo {author} {\bibfnamefont {J.~F.}\ \bibnamefont {Ciparisse}}, \bibinfo
  {author} {\bibfnamefont {R.}~\bibnamefont {Rossi}}, \bibinfo {author}
  {\bibfnamefont {C.}~\bibnamefont {Bellecci}}, \ and\ \bibinfo {author}
  {\bibfnamefont {P.}~\bibnamefont {Gaudio}},\ }\bibfield  {title} {\enquote
  {\bibinfo {title} {{A review of dangerous dust in fusion reactors: From its
  creation to its resuspension in case of LOCA and LOVA}},}\ }\href {\doibase
  10.3390/en9080578} {\bibfield  {journal} {\bibinfo  {journal} {Energies}\
  }\textbf {\bibinfo {volume} {9}} (\bibinfo {year} {2016}),\
  10.3390/en9080578}\BibitemShut {NoStop}%
\bibitem [{\citenamefont {P{\"{u}}tterich}\ \emph {et~al.}(2010)\citenamefont
  {P{\"{u}}tterich}, \citenamefont {Neu}, \citenamefont {Dux}, \citenamefont
  {Whiteford}, \citenamefont {O'Mullane},\ and\ \citenamefont
  {Summers}}]{Putterich2010}%
  \BibitemOpen
  \bibfield  {author} {\bibinfo {author} {\bibfnamefont {T.}~\bibnamefont
  {P{\"{u}}tterich}}, \bibinfo {author} {\bibfnamefont {R.}~\bibnamefont
  {Neu}}, \bibinfo {author} {\bibfnamefont {R.}~\bibnamefont {Dux}}, \bibinfo
  {author} {\bibfnamefont {A.~D.}\ \bibnamefont {Whiteford}}, \bibinfo {author}
  {\bibfnamefont {M.~G.}\ \bibnamefont {O'Mullane}}, \ and\ \bibinfo {author}
  {\bibfnamefont {H.~P.}\ \bibnamefont {Summers}},\ }\bibfield  {title}
  {\enquote {\bibinfo {title} {{Calculation and experimental test of the
  cooling factor of tungsten}},}\ }\href {\doibase
  10.1088/0029-5515/50/2/025012} {\bibfield  {journal} {\bibinfo  {journal}
  {Nuclear Fusion}\ }\textbf {\bibinfo {volume} {50}} (\bibinfo {year}
  {2010}),\ 10.1088/0029-5515/50/2/025012}\BibitemShut {NoStop}%
\bibitem [{\citenamefont {{De Vries}}\ \emph {et~al.}(2011)\citenamefont {{De
  Vries}}, \citenamefont {Johnson}, \citenamefont {Alper}, \citenamefont
  {Buratti}, \citenamefont {Hender}, \citenamefont {Koslowski},\ and\
  \citenamefont {Riccardo}}]{DeVries2011}%
  \BibitemOpen
  \bibfield  {author} {\bibinfo {author} {\bibfnamefont {P.~C.}\ \bibnamefont
  {{De Vries}}}, \bibinfo {author} {\bibfnamefont {M.~F.}\ \bibnamefont
  {Johnson}}, \bibinfo {author} {\bibfnamefont {B.}~\bibnamefont {Alper}},
  \bibinfo {author} {\bibfnamefont {P.}~\bibnamefont {Buratti}}, \bibinfo
  {author} {\bibfnamefont {T.~C.}\ \bibnamefont {Hender}}, \bibinfo {author}
  {\bibfnamefont {H.~R.}\ \bibnamefont {Koslowski}}, \ and\ \bibinfo {author}
  {\bibfnamefont {V.}~\bibnamefont {Riccardo}},\ }\bibfield  {title} {\enquote
  {\bibinfo {title} {{Survey of disruption causes at JET}},}\ }\href {\doibase
  10.1088/0029-5515/51/5/053018} {\bibfield  {journal} {\bibinfo  {journal}
  {Nuclear Fusion}\ }\textbf {\bibinfo {volume} {51}} (\bibinfo {year}
  {2011}),\ 10.1088/0029-5515/51/5/053018}\BibitemShut {NoStop}%
\bibitem [{\citenamefont {Longhurst}\ and\ \citenamefont
  {Snead}(2004)}]{Longhurst2004}%
  \BibitemOpen
  \bibfield  {author} {\bibinfo {author} {\bibfnamefont {G.~R.}\ \bibnamefont
  {Longhurst}}\ and\ \bibinfo {author} {\bibfnamefont {L.~L.}\ \bibnamefont
  {Snead}},\ }\bibfield  {title} {\enquote {\bibinfo {title} {{2 . Status of
  Beryllium Study}},}\ }\href@noop {} {\  (\bibinfo {year} {2004})}\BibitemShut
  {NoStop}%
\bibitem [{\citenamefont {Roth}\ \emph {et~al.}(2008)\citenamefont {Roth},
  \citenamefont {Tsitrone}, \citenamefont {Loarer}, \citenamefont {Philipps},
  \citenamefont {Brezinsek}, \citenamefont {Loarte}, \citenamefont {Counsell},
  \citenamefont {Doerner}, \citenamefont {Schmid}, \citenamefont
  {Ogorodnikova},\ and\ \citenamefont {Causey}}]{Roth2008}%
  \BibitemOpen
  \bibfield  {author} {\bibinfo {author} {\bibfnamefont {J.}~\bibnamefont
  {Roth}}, \bibinfo {author} {\bibfnamefont {E.}~\bibnamefont {Tsitrone}},
  \bibinfo {author} {\bibfnamefont {T.}~\bibnamefont {Loarer}}, \bibinfo
  {author} {\bibfnamefont {V.}~\bibnamefont {Philipps}}, \bibinfo {author}
  {\bibfnamefont {S.}~\bibnamefont {Brezinsek}}, \bibinfo {author}
  {\bibfnamefont {A.}~\bibnamefont {Loarte}}, \bibinfo {author} {\bibfnamefont
  {G.~F.}\ \bibnamefont {Counsell}}, \bibinfo {author} {\bibfnamefont {R.~P.}\
  \bibnamefont {Doerner}}, \bibinfo {author} {\bibfnamefont {K.}~\bibnamefont
  {Schmid}}, \bibinfo {author} {\bibfnamefont {O.~V.}\ \bibnamefont
  {Ogorodnikova}}, \ and\ \bibinfo {author} {\bibfnamefont {R.~A.}\
  \bibnamefont {Causey}},\ }\bibfield  {title} {\enquote {\bibinfo {title}
  {{Tritium inventory in ITER plasma-facing materials and tritium removal
  procedures}},}\ }\href {\doibase 10.1088/0741-3335/50/10/103001} {\bibfield
  {journal} {\bibinfo  {journal} {Plasma Physics and Controlled Fusion}\
  }\textbf {\bibinfo {volume} {50}},\ \bibinfo {pages} {103001} (\bibinfo
  {year} {2008})}\BibitemShut {NoStop}%
\bibitem [{\citenamefont {Allen}(1992)}]{Allen1992}%
  \BibitemOpen
  \bibfield  {author} {\bibinfo {author} {\bibfnamefont {J.~E.}\ \bibnamefont
  {Allen}},\ }\bibfield  {title} {\enquote {\bibinfo {title} {{Probe theory -
  the orbital motion approach}},}\ }\href {\doibase 10.1088/0031-8949/45/5/013}
  {\bibfield  {journal} {\bibinfo  {journal} {Physica Scripta}\ }\textbf
  {\bibinfo {volume} {45}},\ \bibinfo {pages} {497--503} (\bibinfo {year}
  {1992})}\BibitemShut {NoStop}%
\bibitem [{\citenamefont {Willis}\ \emph {et~al.}(2012)\citenamefont {Willis},
  \citenamefont {Coppins}, \citenamefont {Bacharis},\ and\ \citenamefont
  {Allen}}]{Willis2012}%
  \BibitemOpen
  \bibfield  {author} {\bibinfo {author} {\bibfnamefont {C.~T.~N.}\
  \bibnamefont {Willis}}, \bibinfo {author} {\bibfnamefont {M.}~\bibnamefont
  {Coppins}}, \bibinfo {author} {\bibfnamefont {M.}~\bibnamefont {Bacharis}}, \
  and\ \bibinfo {author} {\bibfnamefont {J.~E.}\ \bibnamefont {Allen}},\
  }\bibfield  {title} {\enquote {\bibinfo {title} {{Floating potential of large
  dust grains in a collisionless flowing plasma}},}\ }\href {\doibase
  10.1103/PhysRevE.85.036403} {\bibfield  {journal} {\bibinfo  {journal}
  {Physical Review E - Statistical, Nonlinear, and Soft Matter Physics}\ }
  (\bibinfo {year} {2012}),\ 10.1103/PhysRevE.85.036403}\BibitemShut {NoStop}%
\bibitem [{\citenamefont {Kimura}\ and\ \citenamefont
  {Mann}(1998)}]{Kimura1998}%
  \BibitemOpen
  \bibfield  {author} {\bibinfo {author} {\bibfnamefont {H.}~\bibnamefont
  {Kimura}}\ and\ \bibinfo {author} {\bibfnamefont {I.}~\bibnamefont {Mann}},\
  }\bibfield  {title} {\enquote {\bibinfo {title} {{The Electric Charging of
  Interstellar Dust in the Solar System and Consequences for Its Dynamics}},}\
  }\href {\doibase 10.1086/305613} {\bibfield  {journal} {\bibinfo  {journal}
  {The Astrophysical Journal}\ }\textbf {\bibinfo {volume} {499}},\ \bibinfo
  {pages} {454--462} (\bibinfo {year} {1998})}\BibitemShut {NoStop}%
\bibitem [{\citenamefont {Hutchinson}(2005)}]{Hutchinson2005}%
  \BibitemOpen
  \bibfield  {author} {\bibinfo {author} {\bibfnamefont {I.~H.}\ \bibnamefont
  {Hutchinson}},\ }\bibfield  {title} {\enquote {\bibinfo {title} {{Ion
  collection by a sphere in a flowing plasma: 3. Floating potential and drag
  force}},}\ }\href {\doibase 10.1088/0741-3335/47/1/005} {\bibfield  {journal}
  {\bibinfo  {journal} {Plasma Physics and Controlled Fusion}\ }\textbf
  {\bibinfo {volume} {47}},\ \bibinfo {pages} {71--87} (\bibinfo {year}
  {2005})}\BibitemShut {NoStop}%
\bibitem [{\citenamefont {Delzanno}\ \emph {et~al.}(2013)\citenamefont
  {Delzanno}, \citenamefont {Camporeale}, \citenamefont {{David Moulton}},
  \citenamefont {Borovsky}, \citenamefont {MacDonald},\ and\ \citenamefont
  {Thomsen}}]{Delzanno2013}%
  \BibitemOpen
  \bibfield  {author} {\bibinfo {author} {\bibfnamefont {G.~L.}\ \bibnamefont
  {Delzanno}}, \bibinfo {author} {\bibfnamefont {E.}~\bibnamefont
  {Camporeale}}, \bibinfo {author} {\bibfnamefont {J.}~\bibnamefont {{David
  Moulton}}}, \bibinfo {author} {\bibfnamefont {J.~E.}\ \bibnamefont
  {Borovsky}}, \bibinfo {author} {\bibfnamefont {E.~A.}\ \bibnamefont
  {MacDonald}}, \ and\ \bibinfo {author} {\bibfnamefont {M.~F.}\ \bibnamefont
  {Thomsen}},\ }\bibfield  {title} {\enquote {\bibinfo {title} {{CPIC: A
  curvilinear particle-in-cell code for plasma-material interaction
  studies}},}\ }\href {\doibase 10.1109/TPS.2013.2290060} {\bibfield  {journal}
  {\bibinfo  {journal} {IEEE Transactions on Plasma Science}\ }\textbf
  {\bibinfo {volume} {41}},\ \bibinfo {pages} {3577--3587} (\bibinfo {year}
  {2013})},\ \Eprint {http://arxiv.org/abs/1311.2286} {arXiv:1311.2286}
  \BibitemShut {NoStop}%
\bibitem [{\citenamefont {Rizopoulou}\ \emph {et~al.}(2013)\citenamefont
  {Rizopoulou}, \citenamefont {Robinson}, \citenamefont {Coppins},\ and\
  \citenamefont {Bacharis}}]{Rizopoulou2013}%
  \BibitemOpen
  \bibfield  {author} {\bibinfo {author} {\bibfnamefont {N.}~\bibnamefont
  {Rizopoulou}}, \bibinfo {author} {\bibfnamefont {A.~P.}\ \bibnamefont
  {Robinson}}, \bibinfo {author} {\bibfnamefont {M.}~\bibnamefont {Coppins}}, \
  and\ \bibinfo {author} {\bibfnamefont {M.}~\bibnamefont {Bacharis}},\
  }\bibfield  {title} {\enquote {\bibinfo {title} {{A kinetic study of the
  source-collector sheath system in a drifting plasma}},}\ }\href {\doibase
  10.1088/0963-0252/22/3/035003} {\bibfield  {journal} {\bibinfo  {journal}
  {Plasma Sources Science and Technology}\ }\textbf {\bibinfo {volume} {22}}
  (\bibinfo {year} {2013}),\ 10.1088/0963-0252/22/3/035003}\BibitemShut
  {NoStop}%
\bibitem [{\citenamefont {Bohm}(1949)}]{Bohm1949}%
  \BibitemOpen
  \bibfield  {author} {\bibinfo {author} {\bibfnamefont {D.}~\bibnamefont
  {Bohm}},\ }\href@noop {} {\emph {\bibinfo {title} {{The Characteristics of
  Electrical Discharges in Magnetic Fields}}}},\ edited by\ \bibinfo {editor}
  {\bibfnamefont {A.}~\bibnamefont {Guthrie}}\ and\ \bibinfo {editor}
  {\bibfnamefont {R.~K.}\ \bibnamefont {Wakerling}}\ (\bibinfo  {publisher}
  {McGraw‐Hill Book Company, Inc.},\ \bibinfo {address} {New York},\ \bibinfo
  {year} {1949})\BibitemShut {NoStop}%
\bibitem [{\citenamefont {Bernstein}\ and\ \citenamefont
  {Rabinowitz}(1959)}]{Bernstein1959}%
  \BibitemOpen
  \bibfield  {author} {\bibinfo {author} {\bibfnamefont {I.~B.}\ \bibnamefont
  {Bernstein}}\ and\ \bibinfo {author} {\bibfnamefont {I.~N.}\ \bibnamefont
  {Rabinowitz}},\ }\bibfield  {title} {\enquote {\bibinfo {title} {{Theory of
  electrostatic probes in a low-density plasma}},}\ }\href {\doibase
  10.1063/1.1705900} {\bibfield  {journal} {\bibinfo  {journal} {Physics of
  Fluids}\ }\textbf {\bibinfo {volume} {2}},\ \bibinfo {pages} {112--121}
  (\bibinfo {year} {1959})}\BibitemShut {NoStop}%
\bibitem [{\citenamefont {Laframboise}(1966)}]{Laframboise1966}%
  \BibitemOpen
  \bibfield  {author} {\bibinfo {author} {\bibfnamefont {J.~G.}\ \bibnamefont
  {Laframboise}},\ }\bibfield  {title} {\enquote {\bibinfo {title} {{Theory of
  Sphierical and Cylindrical Langmuir Probes in a Collisionless, Maxwellian
  Plasma at Rest}},}\ }\href@noop {} {\bibfield  {journal} {\bibinfo  {journal}
  {Univ. Toronto Inst. Aerospace Studies Report}\ ,\ \bibinfo {pages} {1--54}}
  (\bibinfo {year} {1966})}\BibitemShut {NoStop}%
\bibitem [{\citenamefont {Pigarov}\ \emph {et~al.}(2005)\citenamefont
  {Pigarov}, \citenamefont {Krasheninnikov}, \citenamefont {Soboleva},\ and\
  \citenamefont {Rognlien}}]{Pigarov2005}%
  \BibitemOpen
  \bibfield  {author} {\bibinfo {author} {\bibfnamefont {A.~Y.}\ \bibnamefont
  {Pigarov}}, \bibinfo {author} {\bibfnamefont {S.~I.}\ \bibnamefont
  {Krasheninnikov}}, \bibinfo {author} {\bibfnamefont {T.~K.}\ \bibnamefont
  {Soboleva}}, \ and\ \bibinfo {author} {\bibfnamefont {T.~D.}\ \bibnamefont
  {Rognlien}},\ }\bibfield  {title} {\enquote {\bibinfo {title} {{Dust-particle
  transport in tokamak edge plasmas}},}\ }\href {\doibase 10.1063/1.2145157}
  {\bibfield  {journal} {\bibinfo  {journal} {Physics of Plasmas}\ }\textbf
  {\bibinfo {volume} {12}},\ \bibinfo {pages} {1--15} (\bibinfo {year}
  {2005})}\BibitemShut {NoStop}%
\bibitem [{\citenamefont {Holgate}\ and\ \citenamefont
  {Coppins}(2018)}]{Holgate2018}%
  \BibitemOpen
  \bibfield  {author} {\bibinfo {author} {\bibfnamefont {J.~T.}\ \bibnamefont
  {Holgate}}\ and\ \bibinfo {author} {\bibfnamefont {M.}~\bibnamefont
  {Coppins}},\ }\bibfield  {title} {\enquote {\bibinfo {title} {{Electron
  emission from electrically isolated spheres}},}\ }\href@noop {} {\ \textbf
  {\bibinfo {volume} {102}} (\bibinfo {year} {2018})}\BibitemShut {NoStop}%
\bibitem [{\citenamefont {Dushman}(1923)}]{Dushman1923}%
  \BibitemOpen
  \bibfield  {author} {\bibinfo {author} {\bibfnamefont {S.}~\bibnamefont
  {Dushman}},\ }\bibfield  {title} {\enquote {\bibinfo {title} {{Electron
  emission from metals}},}\ }\href {\doibase 10.1103/PhysRev.21.623} {\bibfield
   {journal} {\bibinfo  {journal} {Phys. Rev.}\ }\textbf {\bibinfo {volume}
  {21}},\ \bibinfo {pages} {623--636} (\bibinfo {year} {1923})}\BibitemShut
  {NoStop}%
\bibitem [{\citenamefont {Wooldridge}(1957)}]{Wooldridge1957}%
  \BibitemOpen
  \bibfield  {author} {\bibinfo {author} {\bibfnamefont {D.~E.}\ \bibnamefont
  {Wooldridge}},\ }\bibfield  {title} {\enquote {\bibinfo {title} {{Theory of
  secondary emission}},}\ }\href {\doibase 10.1103/PhysRev.56.562} {\bibfield
  {journal} {\bibinfo  {journal} {Physical Review}\ }\textbf {\bibinfo {volume}
  {107}},\ \bibinfo {pages} {977--981} (\bibinfo {year} {1957})}\BibitemShut
  {NoStop}%
\bibitem [{\citenamefont {Thomas}, \citenamefont {Janev},\ and\ \citenamefont
  {Smith}(1992)}]{Thomas1992}%
  \BibitemOpen
  \bibfield  {author} {\bibinfo {author} {\bibfnamefont {E.~W.}\ \bibnamefont
  {Thomas}}, \bibinfo {author} {\bibfnamefont {R.~K.}\ \bibnamefont {Janev}}, \
  and\ \bibinfo {author} {\bibfnamefont {J.}~\bibnamefont {Smith}},\ }\bibfield
   {title} {\enquote {\bibinfo {title} {{Scaling of particle reflection
  coefficients}},}\ }\href {\doibase 10.1016/0168-583X(92)95298-6} {\bibfield
  {journal} {\bibinfo  {journal} {Nuclear Inst. and Methods in Physics
  Research, B}\ }\textbf {\bibinfo {volume} {69}},\ \bibinfo {pages} {427--436}
  (\bibinfo {year} {1992})}\BibitemShut {NoStop}%
\bibitem [{\citenamefont {Schmidt-Ott}\ and\ \citenamefont
  {Federer}(1981)}]{Schmidt-Ott1981}%
  \BibitemOpen
  \bibfield  {author} {\bibinfo {author} {\bibfnamefont {A.}~\bibnamefont
  {Schmidt-Ott}}\ and\ \bibinfo {author} {\bibfnamefont {B.}~\bibnamefont
  {Federer}},\ }\bibfield  {title} {\enquote {\bibinfo {title} {{Photoelectron
  emission from small particles suspended in a gas}},}\ }\href {\doibase
  10.1016/0039-6028(81)90248-X} {\bibfield  {journal} {\bibinfo  {journal}
  {Surface Science}\ }\textbf {\bibinfo {volume} {106}},\ \bibinfo {pages}
  {538--543} (\bibinfo {year} {1981})}\BibitemShut {NoStop}%
\bibitem [{\citenamefont {Rizopoulou}\ and\ \citenamefont
  {Bacharis}(2018)}]{Rizopoulou2018}%
  \BibitemOpen
  \bibfield  {author} {\bibinfo {author} {\bibfnamefont {N.}~\bibnamefont
  {Rizopoulou}}\ and\ \bibinfo {author} {\bibfnamefont {M.}~\bibnamefont
  {Bacharis}},\ }\bibfield  {title} {\enquote {\bibinfo {title} {{Emitting
  large dust grains: Floating potential and potential wells}},}\ }\href
  {\doibase 10.1063/1.5010042} {\bibfield  {journal} {\bibinfo  {journal}
  {Physics of Plasmas}\ }\textbf {\bibinfo {volume} {25}} (\bibinfo {year}
  {2018}),\ 10.1063/1.5010042}\BibitemShut {NoStop}%
\bibitem [{\citenamefont {Krasheninnikov}, \citenamefont {Smirnov},\ and\
  \citenamefont {Rudakov}(2011)}]{Krasheninnikov2011}%
  \BibitemOpen
  \bibfield  {author} {\bibinfo {author} {\bibfnamefont {S.~I.}\ \bibnamefont
  {Krasheninnikov}}, \bibinfo {author} {\bibfnamefont {R.~D.}\ \bibnamefont
  {Smirnov}}, \ and\ \bibinfo {author} {\bibfnamefont {D.~L.}\ \bibnamefont
  {Rudakov}},\ }\bibfield  {title} {\enquote {\bibinfo {title} {{Dust in
  magnetic fusion devices}},}\ }\href {\doibase 10.1088/0741-3335/53/8/083001}
  {\bibfield  {journal} {\bibinfo  {journal} {Plasma Physics and Controlled
  Fusion}\ }\textbf {\bibinfo {volume} {53}},\ \bibinfo {pages} {083001}
  (\bibinfo {year} {2011})}\BibitemShut {NoStop}%
\bibitem [{\citenamefont {Bacharis}\ \emph {et~al.}(2012)\citenamefont
  {Bacharis}, \citenamefont {Coppins}, \citenamefont {Fundamenski},\ and\
  \citenamefont {Allen}}]{Bacharis2012}%
  \BibitemOpen
  \bibfield  {author} {\bibinfo {author} {\bibfnamefont {M.}~\bibnamefont
  {Bacharis}}, \bibinfo {author} {\bibfnamefont {M.}~\bibnamefont {Coppins}},
  \bibinfo {author} {\bibfnamefont {W.}~\bibnamefont {Fundamenski}}, \ and\
  \bibinfo {author} {\bibfnamefont {J.~E.}\ \bibnamefont {Allen}},\ }\bibfield
  {title} {\enquote {\bibinfo {title} {{Modelling of tungsten and beryllium
  dust in ITER}},}\ }\href {\doibase 10.1088/0741-3335/54/8/085010} {\bibfield
  {journal} {\bibinfo  {journal} {Plasma Physics and Controlled Fusion}\
  }\textbf {\bibinfo {volume} {54}},\ \bibinfo {pages} {085010} (\bibinfo
  {year} {2012})}\BibitemShut {NoStop}%
\bibitem [{\citenamefont {Autricque}\ \emph {et~al.}(2017)\citenamefont
  {Autricque}, \citenamefont {Hong}, \citenamefont {Fedorczak}, \citenamefont
  {Son}, \citenamefont {Lee}, \citenamefont {Song}, \citenamefont {Choe},\ and\
  \citenamefont {Grisolia}}]{Autricque2016}%
  \BibitemOpen
  \bibfield  {author} {\bibinfo {author} {\bibfnamefont {A.}~\bibnamefont
  {Autricque}}, \bibinfo {author} {\bibfnamefont {S.~H.}\ \bibnamefont {Hong}},
  \bibinfo {author} {\bibfnamefont {N.}~\bibnamefont {Fedorczak}}, \bibinfo
  {author} {\bibfnamefont {S.~H.}\ \bibnamefont {Son}}, \bibinfo {author}
  {\bibfnamefont {H.~Y.}\ \bibnamefont {Lee}}, \bibinfo {author} {\bibfnamefont
  {I.}~\bibnamefont {Song}}, \bibinfo {author} {\bibfnamefont {W.}~\bibnamefont
  {Choe}}, \ and\ \bibinfo {author} {\bibfnamefont {C.}~\bibnamefont
  {Grisolia}},\ }\bibfield  {title} {\enquote {\bibinfo {title} {{Simulation of
  W dust transport in the KSTAR tokamak, comparison with fast camera data}},}\
  }\href {\doibase 10.1016/j.nme.2016.11.012} {\bibfield  {journal} {\bibinfo
  {journal} {Nuclear Materials and Energy}\ }\textbf {\bibinfo {volume} {12}},\
  \bibinfo {pages} {599--604} (\bibinfo {year} {2017})}\BibitemShut {NoStop}%
\bibitem [{\citenamefont {Tanaka}\ \emph {et~al.}(2007)\citenamefont {Tanaka},
  \citenamefont {Pigarov}, \citenamefont {Smirnov}, \citenamefont
  {Krasheninnikov}, \citenamefont {Ohno},\ and\ \citenamefont
  {Uesugi}}]{Tanaka2007}%
  \BibitemOpen
  \bibfield  {author} {\bibinfo {author} {\bibfnamefont {Y.}~\bibnamefont
  {Tanaka}}, \bibinfo {author} {\bibfnamefont {A.~Y.}\ \bibnamefont {Pigarov}},
  \bibinfo {author} {\bibfnamefont {R.~D.}\ \bibnamefont {Smirnov}}, \bibinfo
  {author} {\bibfnamefont {S.~I.}\ \bibnamefont {Krasheninnikov}}, \bibinfo
  {author} {\bibfnamefont {N.}~\bibnamefont {Ohno}}, \ and\ \bibinfo {author}
  {\bibfnamefont {Y.}~\bibnamefont {Uesugi}},\ }\bibfield  {title} {\enquote
  {\bibinfo {title} {{Modeling of dust-particle behavior for different
  materials in plasmas}},}\ }\href {\doibase 10.1063/1.2722274} {\bibfield
  {journal} {\bibinfo  {journal} {Physics of Plasmas}\ }\textbf {\bibinfo
  {volume} {14}} (\bibinfo {year} {2007}),\ 10.1063/1.2722274}\BibitemShut
  {NoStop}%
\bibitem [{\citenamefont {Gervasini}, \citenamefont {Lazzaro},\ and\
  \citenamefont {Uccello}(2017)}]{Gervasini2017}%
  \BibitemOpen
  \bibfield  {author} {\bibinfo {author} {\bibfnamefont {G.}~\bibnamefont
  {Gervasini}}, \bibinfo {author} {\bibfnamefont {E.}~\bibnamefont {Lazzaro}},
  \ and\ \bibinfo {author} {\bibfnamefont {A.}~\bibnamefont {Uccello}},\
  }\bibfield  {title} {\enquote {\bibinfo {title} {{Physical and Numerical
  Model for Calculation of Ensembles of Trajectories of Dust Particles in a
  Tokamak}},}\ }\href {\doibase 10.1007/s10894-016-0119-5} {\bibfield
  {journal} {\bibinfo  {journal} {Journal of Fusion Energy}\ }\textbf {\bibinfo
  {volume} {36}},\ \bibinfo {pages} {25--39} (\bibinfo {year}
  {2017})}\BibitemShut {NoStop}%
\bibitem [{\citenamefont {Vignitchouk}, \citenamefont {Tolias},\ and\
  \citenamefont {Ratynskaia}(2014)}]{Vignitchouk2014}%
  \BibitemOpen
  \bibfield  {author} {\bibinfo {author} {\bibfnamefont {L.}~\bibnamefont
  {Vignitchouk}}, \bibinfo {author} {\bibfnamefont {P.}~\bibnamefont {Tolias}},
  \ and\ \bibinfo {author} {\bibfnamefont {S.}~\bibnamefont {Ratynskaia}},\
  }\bibfield  {title} {\enquote {\bibinfo {title} {{Dust–wall and
  dust–plasma interaction in the MIGRAINe code}},}\ }\href {\doibase
  10.1088/0741-3335/56/9/095005} {\bibfield  {journal} {\bibinfo  {journal}
  {Plasma Physics and Controlled Fusion}\ }\textbf {\bibinfo {volume} {56}},\
  \bibinfo {pages} {095005} (\bibinfo {year} {2014})}\BibitemShut {NoStop}%
\bibitem [{\citenamefont {Vignitchouk}, \citenamefont {Ratynskaia},\ and\
  \citenamefont {Tolias}(2017)}]{Vignitchouk2017}%
  \BibitemOpen
  \bibfield  {author} {\bibinfo {author} {\bibfnamefont {L.}~\bibnamefont
  {Vignitchouk}}, \bibinfo {author} {\bibfnamefont {S.}~\bibnamefont
  {Ratynskaia}}, \ and\ \bibinfo {author} {\bibfnamefont {P.}~\bibnamefont
  {Tolias}},\ }\bibfield  {title} {\enquote {\bibinfo {title} {{Analytical
  model of particle and heat flux collection by dust immersed in dense
  magnetized plasmas}},}\ }\href {\doibase 10.1088/1361-6587/aa7c44} {\bibfield
   {journal} {\bibinfo  {journal} {Plasma Physics and Controlled Fusion}\
  }\textbf {\bibinfo {volume} {59}} (\bibinfo {year} {2017}),\
  10.1088/1361-6587/aa7c44}\BibitemShut {NoStop}%
\bibitem [{\citenamefont {Laframboise}\ and\ \citenamefont
  {Rubinstein}(1976)}]{Laframboise1976}%
  \BibitemOpen
  \bibfield  {author} {\bibinfo {author} {\bibfnamefont {J.~G.}\ \bibnamefont
  {Laframboise}}\ and\ \bibinfo {author} {\bibfnamefont {J.}~\bibnamefont
  {Rubinstein}},\ }\bibfield  {title} {\enquote {\bibinfo {title} {{Theory of a
  spherical probe in a collisionless magnetoplasma}},}\ }\href {\doibase
  10.1063/1.861425} {\bibfield  {journal} {\bibinfo  {journal} {Physics of
  Fluids}\ }\textbf {\bibinfo {volume} {19}},\ \bibinfo {pages} {1900--1908}
  (\bibinfo {year} {1976})}\BibitemShut {NoStop}%
\bibitem [{\citenamefont {Patacchini}, \citenamefont {Hutchinson},\ and\
  \citenamefont {Lapenta}(2007)}]{Patacchini2007}%
  \BibitemOpen
  \bibfield  {author} {\bibinfo {author} {\bibfnamefont {L.}~\bibnamefont
  {Patacchini}}, \bibinfo {author} {\bibfnamefont {I.~H.}\ \bibnamefont
  {Hutchinson}}, \ and\ \bibinfo {author} {\bibfnamefont {G.}~\bibnamefont
  {Lapenta}},\ }\bibfield  {title} {\enquote {\bibinfo {title} {{Electron
  collection by a negatively charged sphere in a collisionless
  magnetoplasma}},}\ }\href {\doibase 10.1063/1.2741249} {\bibfield  {journal}
  {\bibinfo  {journal} {Physics of Plasmas}\ }\textbf {\bibinfo {volume} {14}}
  (\bibinfo {year} {2007}),\ 10.1063/1.2741249}\BibitemShut {NoStop}%
\bibitem [{\citenamefont {Lange}(2016)}]{Lange2016}%
  \BibitemOpen
  \bibfield  {author} {\bibinfo {author} {\bibfnamefont {D.}~\bibnamefont
  {Lange}},\ }\bibfield  {title} {\enquote {\bibinfo {title} {{Floating surface
  potential of spherical dust grains in magnetized plasmas}},}\ }\href
  {\doibase 10.1017/S0022377815001464} {\bibfield  {journal} {\bibinfo
  {journal} {Journal of Plasma Physics}\ }\textbf {\bibinfo {volume} {82}},\
  \bibinfo {pages} {905820101} (\bibinfo {year} {2016})}\BibitemShut {NoStop}%
\bibitem [{\citenamefont {Thomas}\ and\ \citenamefont
  {Holgate}(2016)}]{Thomas2016}%
  \BibitemOpen
  \bibfield  {author} {\bibinfo {author} {\bibfnamefont {D.~M.}\ \bibnamefont
  {Thomas}}\ and\ \bibinfo {author} {\bibfnamefont {J.~T.}\ \bibnamefont
  {Holgate}},\ }\bibfield  {title} {\enquote {\bibinfo {title} {A treecode to
  simulate dust{\textendash}plasma interactions},}\ }\href {\doibase
  10.1088/1361-6587/59/2/025002} {\bibfield  {journal} {\bibinfo  {journal}
  {Plasma Physics and Controlled Fusion}\ }\textbf {\bibinfo {volume} {59}},\
  \bibinfo {pages} {025002} (\bibinfo {year} {2016})}\BibitemShut {NoStop}%
\bibitem [{\citenamefont {Sonmor}\ and\ \citenamefont
  {Laframboise}(1991)}]{Sonmor1991}%
  \BibitemOpen
  \bibfield  {author} {\bibinfo {author} {\bibfnamefont {L.~J.}\ \bibnamefont
  {Sonmor}}\ and\ \bibinfo {author} {\bibfnamefont {J.~G.}\ \bibnamefont
  {Laframboise}},\ }\bibfield  {title} {\enquote {\bibinfo {title} {{Exact
  current to a spherical electrode in a collisionless, large-Debye-length
  magnetoplasma}},}\ }\href {\doibase 10.1063/1.859619} {\bibfield  {journal}
  {\bibinfo  {journal} {Physics of Fluids B}\ }\textbf {\bibinfo {volume}
  {3}},\ \bibinfo {pages} {2472--2490} (\bibinfo {year} {1991})}\BibitemShut
  {NoStop}%
\bibitem [{\citenamefont {Whipple}(1965)}]{Whipple1965}%
  \BibitemOpen
  \bibfield  {author} {\bibinfo {author} {\bibfnamefont {E.~C.}\ \bibnamefont
  {Whipple}},\ }\emph {\bibinfo {title} {{The equilibrium electric potential of
  a body in the upper atmosphere and in interplanetary space}}},\ \href
  {http://ntrs.nasa.gov/search.jsp?R=19660007937} {Ph.D. thesis},\ \bibinfo
  {school} {George Washington} (\bibinfo {year} {1965})\BibitemShut {NoStop}%
\bibitem [{\citenamefont {Sanmartin}(1970)}]{Sanmartin1970}%
  \BibitemOpen
  \bibfield  {author} {\bibinfo {author} {\bibfnamefont {J.~R.}\ \bibnamefont
  {Sanmartin}},\ }\bibfield  {title} {\enquote {\bibinfo {title} {{Theory of a
  probe in a strong magnetic field}},}\ }\href {\doibase 10.1063/1.1692776}
  {\bibfield  {journal} {\bibinfo  {journal} {Physics of Fluids}\ }\textbf
  {\bibinfo {volume} {13}},\ \bibinfo {pages} {103--116} (\bibinfo {year}
  {1970})}\BibitemShut {NoStop}%
\bibitem [{\citenamefont {Tsytovich}, \citenamefont {Sato},\ and\ \citenamefont
  {Morfill}(2003)}]{Tsytovich2003}%
  \BibitemOpen
  \bibfield  {author} {\bibinfo {author} {\bibfnamefont {V.~N.}\ \bibnamefont
  {Tsytovich}}, \bibinfo {author} {\bibfnamefont {N.}~\bibnamefont {Sato}}, \
  and\ \bibinfo {author} {\bibfnamefont {G.~E.}\ \bibnamefont {Morfill}},\
  }\bibfield  {title} {\enquote {\bibinfo {title} {{Note on the charging and
  spinning of dust particles in complex plasmas in a strong magnetic field}},}\
  }\href {\doibase 10.1088/1367-2630/5/1/343} {\bibfield  {journal} {\bibinfo
  {journal} {New Journal of Physics}\ }\textbf {\bibinfo {volume} {5}}
  (\bibinfo {year} {2003}),\ 10.1088/1367-2630/5/1/343}\BibitemShut {NoStop}%
\bibitem [{\citenamefont {Makkonen}, \citenamefont {Airila},\ and\
  \citenamefont {Kurki-Suonio}(2015)}]{Makkonen2015}%
  \BibitemOpen
  \bibfield  {author} {\bibinfo {author} {\bibfnamefont {T.}~\bibnamefont
  {Makkonen}}, \bibinfo {author} {\bibfnamefont {M.~I.}\ \bibnamefont
  {Airila}}, \ and\ \bibinfo {author} {\bibfnamefont {T.}~\bibnamefont
  {Kurki-Suonio}},\ }\bibfield  {title} {\enquote {\bibinfo {title}
  {{Generating equally weighted test particles from the one-way flux of a
  drifting Maxwellian}},}\ }\href {\doibase 10.1088/0031-8949/90/1/015204}
  {\bibfield  {journal} {\bibinfo  {journal} {Physica Scripta}\ }\textbf
  {\bibinfo {volume} {90}} (\bibinfo {year} {2015}),\
  10.1088/0031-8949/90/1/015204}\BibitemShut {NoStop}%
\bibitem [{\citenamefont {Boris}(1970)}]{Boris1970}%
  \BibitemOpen
  \bibfield  {author} {\bibinfo {author} {\bibfnamefont {J.~P.}\ \bibnamefont
  {Boris}},\ }\bibfield  {title} {\enquote {\bibinfo {title} {{Relativistic
  Plasma Simulation-Optimization}},}\ }\href@noop {} {\bibfield  {journal}
  {\bibinfo  {journal} {4th Conference on Numerical Simulation of Plasma}\ ,\
  \bibinfo {pages} {3}} (\bibinfo {year} {1970})}\BibitemShut {NoStop}%
\bibitem [{\citenamefont {Kennedy}\ and\ \citenamefont
  {Allen}(2003)}]{Kennedy2003}%
  \BibitemOpen
  \bibfield  {author} {\bibinfo {author} {\bibfnamefont {R.~V.}\ \bibnamefont
  {Kennedy}}\ and\ \bibinfo {author} {\bibfnamefont {J.~E.}\ \bibnamefont
  {Allen}},\ }\bibfield  {title} {\enquote {\bibinfo {title} {{The floating
  potential of spherical probes and dust grains . II : Orbital motion
  theory}},}\ }\href {\doibase 10.1017/S0022377803002265} {\bibfield  {journal}
  {\bibinfo  {journal} {Journal of Plasma Physics}\ }\textbf {\bibinfo {volume}
  {69}},\ \bibinfo {pages} {485--506} (\bibinfo {year} {2003})}\BibitemShut
  {NoStop}%
\end{thebibliography}%

\end{document}